\documentclass[conference]{IEEEtran}
\IEEEoverridecommandlockouts

\usepackage{cite}
\usepackage{amsmath,amssymb,amsfonts}
\usepackage{algorithmic}
\usepackage{graphicx}
\usepackage{textcomp}
\usepackage{xcolor}
\usepackage{xspace}
\usepackage[unicode=true,
hidelinks,
pdfauthor={},
pdftitle={Behind the Last Line of Defense --- Surviving Microhypervisor Intrusions},
pdfsubject={}, 
pdfkeywords={},
colorlinks=true,
allcolors=black,
citecolor=blue]{hyperref}

\usepackage[normalem]{ulem}
\usepackage{flushend}

\usepackage{units}
\usepackage{ifthen}
\usepackage{listings}

\newboolean{showcomments}
\setboolean{showcomments}{false}

\ifthenelse{\boolean{showcomments}}
{ \newcommand{\mynote}[3]{
\fbox{\bfseries\sffamily\scriptsize#1}
{\small$\blacktriangleright$\textsf{\emph{\color{#3}{#2}}}$\blacktriangleleft$}}}
{ \newcommand{\mynote}[3]{}}

\definecolor{vrpink}{RGB}{255,0,127}
\definecolor{orange}{RGB}{225,128,0}
\definecolor{brown}{RGB}{225,128,128}
\definecolor{purple}{rgb}{0.54, 0.17, 0.89}

\newcommand{\longo}[1]{}

\usepackage[hide]{chato-notes}

\newcommand{\note}[1]{\nnote{#1}}

\ifthenelse{\boolean{showcomments}}
{ 
}{}

\ifthenelse{\boolean{showcomments}}
{\newcommand{\outline}[1]{\fbox{\begin{minipage}{\linewidth}{\footnotesize#1}\end{minipage}}\\}}{\newcommand{\outline}[1]{}}

\ifthenelse{\boolean{showcomments}}{
  \newcommand{\old}[1]{{\color{blue} Old: #1}}
}{
  \newcommand{\old}[1]{}
}

\newcommand{\midir}{\emph{Midir}\xspace}
\newcommand{\bluedot}{\emph{T2H2}\xspace}

\begin{document}

\title{Behind the Last Line of Defense \\ \Large{Surviving SoC Faults and Intrusions}}

\author{
  \IEEEauthorblockN{In\^es Pinto Gouveia, Marcus V\"olp and Paulo Esteves-Verissimo}
  \IEEEauthorblockA{
    University of Luxembourg\\
    Interdisciplinary Center for Security, Reliability and Trust (SnT) - CritiX group\\
    ines.gouveia@uni.lu, marcus.voelp@uni.lu, paulo.verissimo@uni.lu}
}
\maketitle

\maketitle

\begin{abstract}
  Today, leveraging the enormous modular power, diversity and
  flexibility of manycore systems-on-a-chip (SoCs) requires careful
  orchestration of complex resources, a task left to low-level
  software, e.g. hypervisors.  In current architectures, this software
  forms a single point of failure and worthwhile target for attacks:
  once compromised, adversaries gain access to all information and
  full control over the platform and the environment it controls.
  This paper proposes \midir, an enhanced manycore architecture,
  effecting a paradigm shift from SoCs to distributed SoCs.  \midir
  changes the way platform resources are controlled, by retrofitting   
  tile-based fault containment
  through well known mechanisms, while securing
  low-overhead quorum-based consensus on all critical operations,
  in particular privilege management and, thus, management of
  containment domains.
  Allowing versatile redundancy management, \midir promotes resilience
  for all software levels, including at low level.  We explain this
  architecture, its associated algorithms and hardware mechanisms and
  show, for the example of a Byzantine fault tolerant microhypervisor,
  that it outperforms the highly efficient MinBFT by one order of
  magnitude.

\end{abstract}


\section{Introduction}
\label{sec:introduction}

Practically all activity of modern societies depends on
information and communication technologies (ICT).  Such dependency
obviously hinges on the correctness of these systems, some of them
critical, which may fail in a combination of multiple causes and
ways~\cite{facebook_privacy,singapore_health_privacy,under_armour,CloudOutage,ukrain_power,tesla}.
Systems have been progressively pushed to extremes of efficiency
through modularity in platform sharing, firstly through virtualization
and lately by
leveraging the enormous power
growth, functional diversity and adaptation flexibility offered by
multi- and manycore.
This has taken platform sharing to new heights,
into the realm of multi-processor systems-on-a-chip (MPSoCs).

The organization of these complex computing resources depends on
low-level platform management hardware (e.g., memory-management units
(MMUs)) and software (e.g., firmware, hypervisors, management
engines). However, current MPSoC architectures are such that these
management components, which should form a last line of defense
against severe accidental faults or adversaries intruding the system (malicious faults), instead
constitute a single point of failure (\emph{SPoF}), for two main
reasons. First, the way platform privilege-enforcement mechanisms
(e.g. MMUs or hardware-enforced
capabilities~\cite{Woodruff:2014:CCM:2665671.2665740}) are designed
allows faults in a core/tile to propagate through MPSoC components.
Second, faults in this lowest-level management software, e.g.,
hypervisors configuring these privileges, are bound to propagate
across management and managed components, again causing common-mode
failure scenarios.

If these SPoFs are compromised by adversaries, the latter gain full
authority over the platform's privilege-enforcement mechanisms and,
through them, access to all information and complete control over all
platform resources (e.g., cloud-based systems), including, in the case
of cyber-physical systems, extended control over the physical
environments on which they act (e.g., nuclear power plants
or autonomous cars).

Is this a real risk? It is, if the vulnerability rate of these
low-level platforms is non-negligible.
Recent problems, whether in Intel's 
CSME~\cite{me_vulnerability}, Xen/Critix~\cite{xen_vulnerabilities}
or concerning
Spectre~\cite{spectre} and Meltdown~\cite{meltdown}, have been
repeatedly reminding us of how brittle the assumption of
``tamperproof and unattackable low-level platform management assets'' is.
Even formally verified kernels (e.g.,
seL4~\cite{Klein+al:sosp:sel4:2009}) may fail due to model/reality
discrepancies or
hardware faults violating modeling assumptions~\cite{Biggs_LH_18}.

Being the risk real, are there no solutions yet?
The solution design space for contemporary hardware platforms
dependability and security has been unfolding in two directions:
(i) application-specific system-level replication (e.g., triple
modular redundancy, mainly in cyber-physical systems (CPS), by means
of multiple electronic control units (ECUs)), where the lack of flexibility limits the extension
to general systems;
(ii) manycore-level replica management and consolidation, which then,
if on bare MPSoCs, reintroduces the SPoF concern, now for the
low-level replication management component.

\nnote{
The design space for contemporary hardware platforms unfolds in two
directions:
(i) System-level replication (e.g., triple modular redundancy by means of
multiple electronic control units (ECUs)), which is limited in flexibility by the number of systems
provisioned. The maximum replication degree $n$ is fixed and, with the
whole manycore system being the fault containment domain, applications
with a smaller replication degree suffer from low-level management
failing simultaneously. Much of the potential is lost that can be
gained from consolidating replicas in the manycore system.
(ii) Consolidation, on the other hand, reintroduces the concern of a single
low-level management component (configuring privilege enforcement)
failing and this fault propagating as a common-mode fault to all
replicas.
A traditional strategy for coping with the latter, potentially
vulnerable software, has been to introduce an underlying, 
assumed-trustworthy layer, charged with locally detecting and confining
errors~\cite{5504713,Seshadri:2007:STH:1294261.1294294} or even
performing fault and intrusion
tolerance~\cite{Chun:2008:DRS:1404014.1404038,romain}.
%

It so happens that these approaches, even if they are ``simple'', 
normally have a relative complexity which is at least at the level of the very
components we seek to protect, like
microkernels~\cite{Liedtke:sosp:l4:1995}, if not quite more.  Even a
residual fault or vulnerability rate in these supposedly trusted
components will defeat our purpose and breach the platform's safety and security. In fact, ``simple'' components
with at least a few KLOCs have a non-negligible statistical fault
footprint~\cite{univis91357728}, not to mention the complexity of the
hardware components that are necessary to execute this code.
}

At this time, we call the reader's attention to an interesting fact,
which will become crucial to our solution. The current MPSoC
architectures' complexity, modularity and networked interconnectivity,
suggests attributes of distributed systems~\cite{SapeMullenderDisSys},
albeit imperfect such systems (an example of which is the
aforementioned SPoF syndrome).  However, distributed systems have been
used to mitigate SPoF syndromes and to implement fault and intrusion
tolerance schemes~\cite{Powell:Delta4,MaftiaSecPrivMag}.  In
consequence, the root of the MPSoC problems just presented may also be
an avenue to their solution.

\note{
At this time, we call the reader's attention to an interesting fact, which
will become crucial to our solution: the root of the MPSoC problems just
presented may also be an avenue to their solution.

In fact, MPSoC complexity, modularity and network interconnection, suggest
attributes of distributed systems~\cite{SapeMullenderDisSys}, which have
been used to mitigate SPoF
syndromes and implement fault and intrusion
tolerance schemes~\cite{Powell:Delta4,MaftiaSecPrivMag}.
However, current MPSoC architectures are imperfect distributed systems.
}

So, in this paper, we start by identifying the gaps from (MP)SoCs to
distributed systems, and
propose (MP)SoC mechanisms to bridge them, which
essentially means achieving: fault independence and fault containment, despite low software-level compromise and while retaining the flexibility (MP)SoCs offer.
Having a manycore that behaves as a (closely-coupled) distributed
system, should allow us to design a set of efficient and low-overhead
distributed systems-inspired modular protection and redundancy
management mechanisms, e.g., Byzantine fault tolerant state machine
replication (BFT-SMR), for fault and intrusion tolerance (FIT).
The remaining problem, how to implement and where to locate all the
mechanisms above, is addressed by the \midir\footnote{pronounced
  meedir} achitecture presented in this paper, which leverages the
computing critical mass and flexibility of contemporary tile-based
manycore architectures.

\midir constrains the connection of all tiles to the network-on-chip
(NoC) through simple and self-contained hardware-based trusted
components, which we call \bluedot.
Exploring the concept of architectural
hybridization~\cite{Verissimo:2006:TTW:1122480.1122497}, whilst we
consider those components to be ultra-reliable and not fail, we are
agnostic about the reliability of individual tiles, which may be
compromised or fail. The assumption is justified by the simplicity of
the former, promoting verifiability.

The \bluedot components implement the functionality achieving
fault independence, containment, and tolerance mechanisms mentioned
above. In consequence, tile-internal software or hardware faults are
contained in the tile and the objects the tile can access.
Furthermore, the baseline mechanisms for protection and redundancy
management provided by \bluedot can be extended and recursively
applied at any software layer, giving the designer ample latitude for
crafting resilience into systems, both ``horizontally'' (incremental
power of defense mechanisms) and ``vertically'' (depth of defense).

Locating \bluedot between the tile and the NoC interconnect not
  only provides a clear pathway for integration by chip manufacturers
  and integrators, it also allows drawing from many well-understood
  building blocks (e.g., region protection, capabilities~\cite{needham:cap},
  and other chip-level resource management
  mechanisms~\cite{config_isolation}, capable of isolating tiles and
  the resources they can access). The novelty of \midir lies in their
  arrangement to avoid SPoFs, even while they are reconfigured.

\note{
Leveraging
the concept of
hybridization~\cite{Verissimo:2006:TTW:1122480.1122497}, we implement
these mechanisms
with the assistance of simple trusted-trustworthy hardware-based components
to achieve ultra-reliable constructs mitigating the residual faults
syndrome depicted earlier.

This baseline structure is materialized by the \midir\footnote{pronounced meedir} architecture, which we present in
this paper. \midir leverages the natural quantitative and
qualitative computing critical mass available in contemporary tile-based
manycore architectures, applying concepts from distributed systems to
mitigate faults and intrusions,
promoting incremental tile-granular resilience for all software
levels, down to the lowest level.

By constraining tile-external access, interposed by the
trusted-trustworthy hardware component, we constrain how software in this
tile and its hardware components can interact with other tiles and
with central hardware. Consequently, faults in this software or hardware
are contained
to the tile and the objects the tile can access.
The reader will note that \midir functionality can be
recursively applied at any software layer, giving the designer ample 
latitude for crafting resilience into
systems, both ``horizontally'' (power of mechanisms) and
``vertically'' (depth of protection).
}

\note{
What now remains to be seen --- the main contribution of this paper
--- is how \midir enables the construction of such a FIT (fault and intrusion tolerant) 
kernel without reintroducing SPoF syndromes when configuring application and
kernel-level privileges.
}

In a nutshell, contributions of this paper are:

(1) An analysis of the gaps separating current MPSoC architectures
from genuine
distributed systems, and gap fixing through measures promoting fault
independence and fault containment in tile-based architectures,
enforced at the level of the tile-to-NoC
interface.

(2) An architecture (\midir) leveraging the resulting distributed
system-on-a-chip (DSoC) to achieve incremental levels of modular fault
and intrusion tolerance, through a range of diverse redundancy
management techniques 
implemented by simple hardware-based voting/consensus mechanisms.

(3) The design of a simple and ultimately {\bf t}rusted-{\bf
    t}rustworthy {\bf h}ardware {\bf h}ybrid, \bluedot --- the core
  component of \midir, staged at the tile-to-NoC interface ---
  providing just two generic baseline functions: access control
  (capability registers) and quorum-based consensus (voters).
  By configurations and combinations of these two basic
  functions, \bluedot is capable of implementing all the techniques
  mentioned in (1) and (2).


(4) As a proof of concept, we give and evaluate an implementation
  featuring \midir and essential parts of a fault and intrusion
  tolerant microhypervisor built on top of it.  Though the
  architecture serves several reliability strategies, we chose the
  most effective, active replication with error masking. Being the
  most complex and costlier, we believe to have shown the performance
  and practicality of our concept.

  %

Next, we evaluate the challenges for bridging from SoCs to DSoCs
(Sec.\ref{sec:manycore}), and present the system and threat model
(Sec.\ref{sec:threat_model}).  Then, we introduce the \midir
architecture (Sec.\ref{sec:muenchhausen}) and the \bluedot component
in Sec.\ref{sec:T2H2}.
At this point, we are able to show \midir in action, discussing the
design of a fault and intrusion tolerant microhypervisor built on top
of it (Sec.\ref{sec:hypervisor}), as an example of critical low-level
management software.
Finally, we discuss some relevant implementation matters in
Sec.~\ref{sec:implementation}, and in Sec.\ref{sec:evaluation}, we
evaluate \midir on a Zynq ZC702 board, showing how \midir's hardware
voters accelerate BFT-SMR protocols, voted execution of system calls
and consensual reconfiguration of \bluedot.  
An analysis of related work (Sec.~\ref{sec:related-work}) follows, and
Sec.\ref{sec:conclusions} concludes the paper, pointing to further
research and innovation opportunities.

\note{
In a nutshell, contributions of this paper are:

(1) An analysis of the gaps separating current MPSoC architectures
from genuine (though closely-coupled) distributed systems, and gap
fixing through measures promoting fault independence and fault
containment: (i) resorting to tile-based architectures;
(ii) enforcing baseline per-tile (``node'')
access control and configuration
  protection,
at the level of the 
hardware components
implementing the tile-to-NoC (network-on-chip) interface; (iii)
protecting critical platform management operations, by imposing
consensus of a majority of correct components. Our findings may be
useful to enhance MPSoC architectures in general, towards distributed
systems-on-a-chip (DSoC).

(3) An architecture (\midir) leveraging the resulting distributed
  system-on-a-chip (DSoC) to achieve incremental levels of modular
  fault and intrusion tolerance, through a range of diverse redundancy
  management techniques (from simple error detection and self-checking
  by comparison, to error masking by voting), implemented by simple
  hardware-based voting/consensus mechanisms.

(4) The design of \bluedot --- the core component of \midir, staged at
the tile-to-NoC interface --- a simple and ultimately {\bf
  t}rusted-{\bf t}rustworthy {\bf h}ardware {\bf h}ybrid, providing
just two generic functions: access control (capability registers) and
quorum-based consensus (voters).  By configurations and combinations
of these two basic functions, we offer efficient, robust and
low-overhead mechanisms and algorithms implementing all the techniques
described in (2) and (3) above.

(5) As a proof of concept, we give and evaluate an implementation featuring \midir and
  essential parts of
  a fault and intrusion tolerant microhypervisor built on top of it.
  Though the architecture serves several incremental reliability
  strategies, we chose the most effective, active replication with
  error masking. Being the most complex and costlier, we believe
  to have shown the practicality of our concept, as having quite
  satisfying performance.
  %

\midir and \bluedot were intentionally designed as non-intrusive
extensions to current chip architectures, allowing a backward
compatible, non-fracturing evolution.  Taken up by a hardware
manufacturer or integrator, they can easily lead to next-generation
COTS resilient chips.

After discussing related work,
we evaluate the challenges for
bridging from SoCs to DSoCs (Sec.\ref{sec:manycore}),
present
the system and threat model (Sec.\ref{sec:threat_model}) and
introduce the \midir architecture in general
(Sec.\ref{sec:muenchhausen}) and \bluedot in Sec.\ref{sec:T2H2}.

We discuss in Sec.\ref{sec:privilege_reversion} the
importance of consensual privilege management 
as a last but fundamental step in the SoC to DSoC metamorphosis in
order to prevent fault propagation and common-mode failures in the
platform.

In Sec.\ref{sec:hypervisor}, we focus on the very important
objective of mitigating the risk of faults and intrusions in the
hypervisor, as an example of critical low-level management software.
%
In Sec.\ref{sec:evaluation}, we evaluate \midir on a Zynq ZC702 board, 
showing how \midir's hardware voters accelerate BFT-SMR protocols,
voted execution of system calls and consensual reconfiguration of \bluedot. 
%
Sec.\ref{sec:conclusions} concludes. 
}

\section{From MPSoCs to Distributed SoCs}
\label{sec:manycore}

Multi- and manycore systems consolidate in a single chip computing
resources that used to reside on multiple chips. Tiles~\cite{raw} are
placeholders and instantiation points for resources, typically
instantiated with cores and private caches or with slices of shared
caches, and connected through the NoC with each other and with memory
controllers (to reach out to RAM/IO). It is possible as well to cast accelerators,
GPUs and FPGAs, into the tile abstraction.

The modularity and networked interconnection of tiles already suggests
attributes of a distributed system and has inspired first steps to
hardware-enforced fault containment at tile level, as pioneered by
Hive~\cite{hive}, Cap~\cite{needham:cap},  M3~\cite{asmussen:m3} and others.
%
Hive introduces MAGIC, a bus-level firewall to confine faults to the
individual processors of the Stanford Flash multiprocessor system. M3
follows the same scheme with hardware enforced capabilities,
originally introduced in Cap~\cite{needham:cap} to control resource
accesses and, thereby, fault containment of heterogeneous processors.
Configurable isolation~\cite{config_isolation} leverages dual-mode
redundant MMUs to, like M3, confine faults in on-chip resources.
%
Tiles favour functional and non-functional diversity since they can
host cores from several makers. This improves fault independence
through the implied low likelihood of experiencing the same fault in
different tiles. Similarly, different versions of the same code can be 
used at distinct tiles with the same intent~\cite{avizienis1977implementation, knight1986experimental, joseph1988fault}.

\note{
Even a large class of hardware-level faults can be confined in this
way, although not all. Conventional multi- and manycore designs retain
the possibility of common mode failures in central hardware components
(e.g., the clock or power distribution network), which must be
addressed differently (e.g., with resilient
clocks~\cite{steininger:clock}). However, as long as physical effects
of a fault are retained to the causing tile and the signals it
exhibits to the system, such faults can be contained through
tile-level privilege enforcement.
}

Note that, emulating the spacial isolation of distributed system
nodes, we are agnostic about the semantics and interplay of
tile-internal and/or core-level components, e.g., MMUs and their
virtualization, copy-on-write, memory protection or recovery
functionalities.

\note{
Note that existing core-level mechanisms, like MMUs, aside from
isolation, have other functions, such as creating the illusion of a
dedicated ``virtual'' memory space and managing optimizations,
such as copy on write. 
We enforce tile-level protection (at the tile-to-NoC interface) for
fault containment since it is what emulates best the spacial
isolation of distributed system nodes.  We aim to be the least intrusive
and so we leave virtualization and recovery
from privilege violations (segfaults, page faults, etc.) to core-level
mechanisms. This strategy has the added benefit of simplifying
\bluedot
and justifying its ultimate trustworthiness
assumption.
}

A final and subtle gap concerning fault containment and independence
affects all previous systems we know of, including those deploying
hardware-enforced fault
containment~\cite{needham:cap,hive,asmussen:m3,config_isolation}: potentially faulty or
compromised low-level kernels still retain control over platform
privilege configuration mechanisms.
As we explain in Sec.\ref{sec:muenchhausen}, this is a harmful effect.
Our main contribution is to neutralize this effect by imposing that critical platform management
operations are performed through consensus of a majority of correct
components.

In conclusion, with the enhancements described in this paper, tiles
fail like nodes in a distributed system, faults affect only the tile
itself and the components (e.g., replicas) executing on it, but they
do not propagate to the entire manycore, in particular other
components related to the same application or subsystem.
This interplay between protection and consensus to achieve fault
  containment, in particular during platform reconfiguration,
  including of the fault containment domains themselves, allows hypervisor
  replicas to retain the flexibility of the MPSoC, even after a
  minority of hypervisor tiles failed accidentally or have been
  compromised by an adversary.

\section{System and Threat Model}
\label{sec:threat_model}

We now describe the system and threat model educating the development of
our distributed system-on-a-chip (DSoC).


\subsection{System Model}

We assume a fully connected system, where on-chip network components
offer the abstraction of a correct network, interconnecting all tiles
to one another. Tiles communicate by messages, and messages sent are
eventually delivered, unchanged, to the destination.
Network coding~\cite{network-coding}, multi-tenant~\cite{multi-tenand}
and adaptive routing techniques~\cite{adaptive-noc-routing}
substantiate the coverage of this assumption.

We rely on a partially-synchronous model. At first sight, manycores
might seem the perfect example of a (closely-coupled) synchronous
(distributed) system. However, reality is a bit different, several
possibilities for instability in the time domain (speed of tiles
throttling for thermal control, cache exceptions, NoC-level bursts,
etc.) would prove the strict synchronous model brittle.

However, being a closely-coupled environment, short-term liveness is
normally guaranteed, barring delay variations.
This has two implications on the design of \midir, for robustness: (i)
we absorb possible inter-tile delays, notably by buffering messages
(e.g., votes) in \midir's \bluedot; (ii) the structure of the
protocols is time-free and, as such, they remain safe in the presence
of delay oscillations, provided that the fault assumptions hold.

\note{
Although manycores might seem the perfect example of a
(closely-coupled) synchronous (distributed) system, reality is a bit
different. There are several possibilities for instability in
the time domain. For example, excessive resource use raises the
temperature and causes thermal managers to throttle the speed of tiles
near this hot spot; interfering access patterns reduce memory
bandwidth by evicting cachelines from shared caches; and NoC-level
bursts may cause noticeable and, with unfair arbitration, potentially unbounded
message delays. Faulty behavior (accidental or malicious) might
further worsen these negative time-domain effects.  A strict
synchronous model would not reflect reality and thus be proved
brittle.

We rely on a partially-synchronous model and prepare \midir for
possible delays (notably by buffering consensus votes in \midir's \bluedot). 
Two particularities exist in these closely-coupled
environments, in contrast to large-scale distributed systems, which
play in our favour: (i) barring delay variations, liveness is normally
guaranteed; and (ii) the infrastructure is plastic in terms of timeliness
trade-offs. Therefore, as in most contemporary BFT approaches, we consider 
asynchrony for safety and partial synchrony for liveness.

The structure of our protocols is time-free and, as such, they remain
safe in the presence of delay oscillations, provided that the
fault assumptions hold (no more than $f$ tiles get compromised, as
discussed below). 
Then, the protocols inherit synchrony from the timeliness of the
infrastructure they are immersed
in~\cite{VerissimoSRDS09,LeLannxxxAskMe}:
the manycore works with high performance, in
execution and communication, exhibiting short and bounded delays
during long enough periods of time, but can exhibit significant
variations in these bounds. These are fair expectations, considering 
the nature of these systems.
}


\subsection{Threat Model}

Our threat model considers software-level compromise at all levels,
including hypervisors, firmware and, more generally, in
any critical software component.  This assumption is consistent with
our aim of tolerating an incremental level of threat on tiled manycore
systems, up to sophisticated and persistent attacks possibly deployed
entirely on-chip.
%
%
Moreover, we consider a limited set of
hardware-level faults and attacks: precisely those whose physical
effects are confined to a tile (e.g., trapdoors in a core,
but no hardware faults that cause a chip-wide collapse). 

We consider the tile as a unit of component failure. There is no 
guaranteed fault containment inside tiles. 
That is, adversaries (or accidents) will be
capable of compromising the whole software in any tile (e.g., but not
only, a hypervisor
in case the user/supervisor mode isolation failed).
Once that
happens, we no longer make any assumptions about the correctness of
any software in that tile. 
However, we also consider (and enforce it with the strategy described
in Sec.\ref{sec:manycore}) that tiles themselves are fault containment
domains, such that faults inside a tile do not propagate across the manycore.  
%
%



We enforce the assumption above through architectural
hybridization~\cite{Verissimo:2006:TTW:1122480.1122497,minbft,cheapbft}.
Despite the general system fault model enunciated for tiles, \bluedot
(\midir's trusted-trustworthy component) follows a more restricted
fault model, enforced by construction and, through its simplicity, amenable to verification, failing only
by crashing, much like USIG~\cite{minbft} or CASH~\cite{cheapbft}. Thus,
\bluedot, residing at the tile-to-NoC interface, reliably implements
its functions despite faulty tiles.


\begin{figure}
\begin{center}
\includegraphics[width=.8\columnwidth]{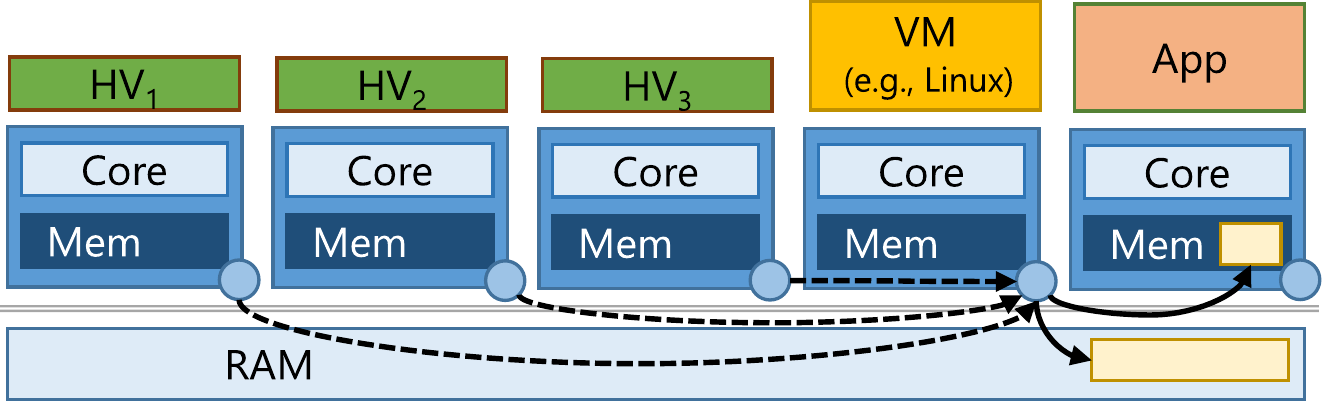}
\caption{Example systems showing the \midir architecture: 
software in tiles need a capability to authorize access to resources
in other tiles (solid lines);
  capability modifications in a tile (in fact any critical
    operation) are subject to consensus of a majority of other
  correct tiles (dashed lines), here the three tiles hosting
  hypervisor replicas of which one may be faulty.  }
\label{fig:architecture}
\end{center}
\end{figure}

\note{
Overview of the \midir architecture: a multi-/manycore
  system augmented with \bluedot hardware units (blue dots) at the NoC
  interface.
  Tiles and the software they execute are able to access only those
  external resources to which they hold a capability, explicitly
  authorizing this access (solid lines on the right).
  Privilege change in a tile (i.e., capability modifications) is
  subject to consensus of a majority of other correct tiles (dashed
  lines on the left).  
}

\section{The \midir Architecture}
\label{sec:muenchhausen}

As discussed earlier, \midir is an architectural concept based
on augmenting manycore systems in a minimally intrusive way through
strategically placed, simple and self-contained trusted-trustworthy
components (\bluedot).
%
%
In fact, \bluedot provides just two generic baseline functions staged
\emph{in hardware} at the tile-to-NoC interface: access control
(capability registers) and quorum-based consensus (voters).

%
%
\old{These are integrated in the \bluedot hardware units (the ``blue
dots'' of Fig.\ref{fig:architecture}), featuring essentially two
functions: capability registers (including the logic for checking
privileges when invoking capabilities) and voters.}%
%

Fig.\ref{fig:architecture} depicts one possible layout, of a
stereotypical hypervisor-based system, where the hypervisor is
replicated for fault/intrusion tolerance, serving operating system and
applications: hypervisor replicas are distributed across tiles, so
that each replica executes on a different tile, separate from
applications; tiles and software therein interface with each other
through the NoC; and \bluedot are the ``blue dots'' performing that
interconnection. \bluedot interposes such accesses, validating that
the invoking tile has sufficient privileges, through the capability
registers, which include the logic for privilege enforcement.

As long as the execution in a tile remains within the resources
associated to this tile (local caches, memories, accelerators, etc.)
no overhead occurs, since \bluedot is not involved in authorizing or
denying these accesses. In fact, we remind that it is not the purpose
of \midir to provide fault containment between software components
co-located {\em on the same tile}.  This is like the internal behavior
of nodes in a distributed system, where nodes are the unit of fault
containment.
%
%
Once software components are spread across tiles, they
interact through external operations (e.g., via a resource in another
tile, via shared on-chip memories
or via external memory or IO) and \bluedot validates that each such
access has
been authorized by a capability the tile possesses.
Consequently, hardware faults inside a tile or accidential or malicious faults in any part of the software it executes, are limited in propagation to the objects authorized by these capabilities.

\note{
Central to \midir is that \bluedot's capability registers are always
considered internal for invocation, but external for
reconfiguration. As such, the latter is always subject to capability
checks and, as we shall see, voting.
}

Further to capability checking, \midir is capable of subjecting these
accesses to voting by distributed components in different tiles. This
is especially important for critical operations, be it in application
execution or in platform reconfiguration, in order to achieve some
form of fault/intrusion tolerance, from error detection, or
self-checking by comparison, to error masking by consensus.
To vote, tiles must hold a capability to the corresponding voter,
  which authorizes this tile to make proposals as one of these
  distributed components. Voting is mandatory to install new or
  change existing capabilities, in order to prevent faulty hypervisor
  replicas from bypassing the aforementioned fault containment when
  reconfiguring the resources a tile can access.

\note{
The key objective of \midir is to make all privilege changes and
likewise all critical operations (e.g., platform reconfiguration or
actuator accesses)
effective only after a fault-tolerant quorum of replicas has reached
consensus to execute the operation (recalling, what we name
'consensual' for simplicity).

As such, we implement capabilities for access control
to tile-external resources (including the
reconfiguration interface of the tile's capability unit) and voters to
interpose
critical operations and ensure consensus before
execution.
}

\midir's concept of controlling the tiles' lowest-level privilege
enforcement mechanism is agnostic of the mechanism used.  However, the
simpler such a mechanism and the closer it can be implemented to the
tile's NoC interconnect, the more architecture-level faults \midir will be
able to tolerate. Hence our choice for capabilities. 

Simplicity also governs our voter design. \midir's voters merely
collect and act upon proposals of related operations from different
components, letting the voted-upon operation proceed.  Because
tile-external resources are typically memory mapped, these operations
are normally simple writes.  The voters themselves implement no error
handling or diagnostics functionality, but provide information for the
components to perform these tasks.
More precisely, voters suspend voting on disagreement, freeze the
proposals made by the components and expose them for
diagnosis. Moreover, they implement a sequence number $\mathit{seq_i}$
for progress tracking, which they increment after each vote unless the
vote gets suspended. A voted upon voter-reset operation resumes voting
and as well increments $\mathit{seq_i}$. Sec.\ref{sec:hypervisor} shows how we
utilize this error handling support and Sec.\ref{sec:implementation}
details our voter implementations.

\section{\bluedot -- Midir's Trusted-Trustworthy Component}
\label{sec:T2H2}

\note{
Before diving into the consequences of consensual privilege change,
which then leads us to the design of \midir-aware FIT hypervisors, let
us provide further details about \bluedot.
}

In this section, we provide further details about \bluedot.

\subsection{Voted and non-voted operations}
\label{sec:direct_invocation}
\label{sec:voted_invocation}

\begin{figure}
\begin{center}
  \includegraphics[width=\columnwidth]{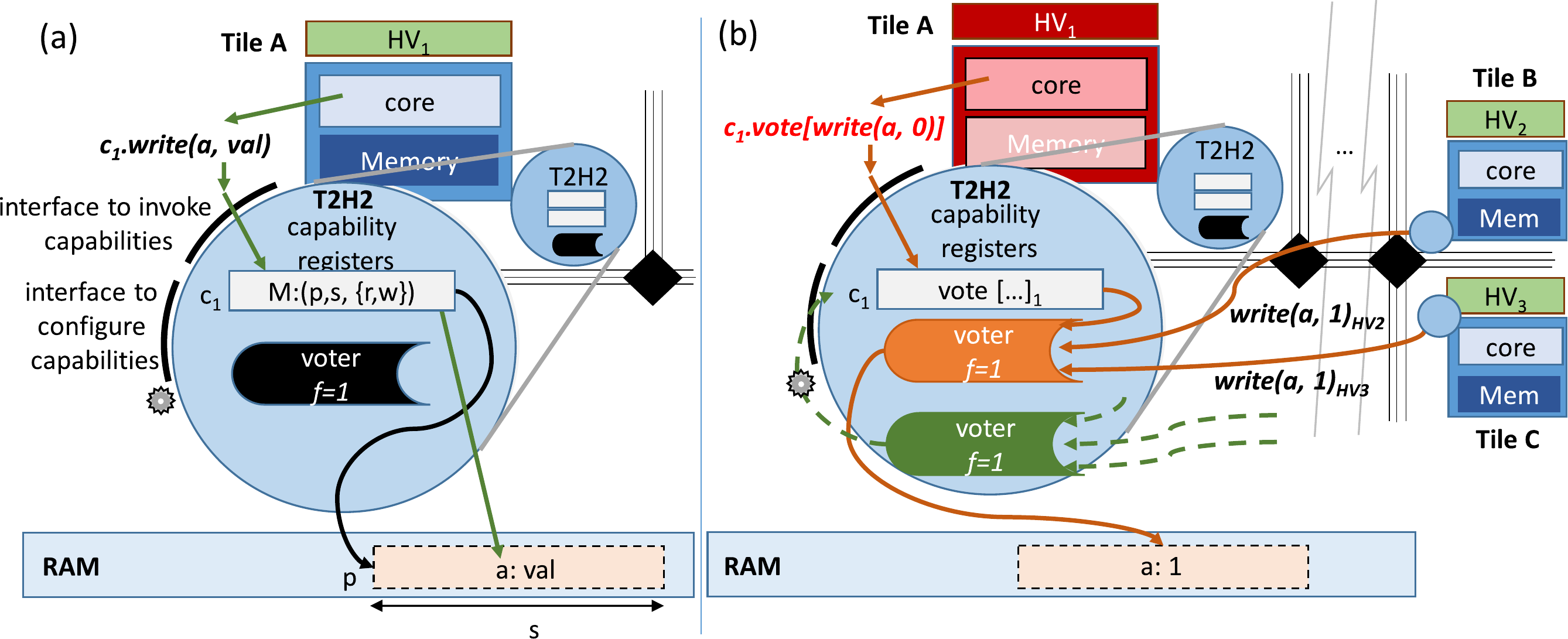}
  \caption{ 
(a) Non-voted memory access by tile A through capability invocation.
(b) Voted memory access by tiles A-C (tile A faulty) through
    capability invocation then voting (orange); reconfiguration of a
    platform capability register in tile A through voting (green).
}
  \label{fig:invocation}
  \vspace{-1mm}
  \end{center}
\end{figure}

To retain the flexibility of the software in a manycore system, allowing 
it to dynamically adapt resource-to-application mappings as needed,
\bluedot supports 
direct 
access to tile-external
resources. This way, applications possessing a capability can directly
invoke operations on external resources (e.g., to access read-shared
or private data in RAM or to interact with non-critical devices).
%
%
The scenario in Fig.\ref{fig:invocation}(a) illustrates a non-voted
(write) memory access by Tile A, performed by invoking a capability in
this tile's \bluedot. Since \bluedot's capability registers hold a
read-write capability to the memory region $[p, p+s]$, the operation
to write value $val$ in variable $a$ is
authorized.

However, \bluedot also supports voting, particularly useful when e.g.,
platform management software or hypervisor replicas,
further have to execute critical operations (e.g.,
privilege change or critical device accesses). 
These operations are voted upon, within preconfigured detection or
tolerance mechanisms, to prevent compromised components from causing
harm.
%
%
Several strategies may be served by \midir, such as self-checking,
recovery blocks, or {\em f-out-of-n} error masking by majority voting
in the presence of $f$ faulty components, but they are all supported
by the same baseline voting mechanism.
Fig.\ref{fig:invocation}(b) represents a similar operation as in
Fig.\ref{fig:invocation}(a), but in voted access form. Tiles B
and C vote to write value 1, while Tile A, being faulty, votes to
write value 0. In order to perform these votes, all tiles invoke a
capability on their local \bluedot to access the designated voter (in
this case, residing on Tile A's \bluedot) . Given that a majority of
tiles voted to write 1, value 1 will be written to variable $a$.

\midir does not constrain how systems are configured and hence
  what faults are tolerated. Instead it provides the means to tolerate
  an incremental quality of faults, including for highly critical
  systems up to $f$ faults in system management software (e.g., the
  hypervisor), by providing $n = 2f+1$ hypervisor replicas and by
  subjecting all critical operations to voting.

\subsection{Consensual privilege change}
\label{sec:privilege_reversion}

One particularly
relevant scenario for voted access is consensual
reconfiguration of the \bluedot instances themselves. \bluedot's
reconfiguration interface is accessible only through a voter and
cannot ever be invoked directly.

Let us understand why this is a relevant innovation.  In conventional
OS design, any single kernel instance can directly or indirectly
enforce modifications on platform resources. So, even in fault
tolerant designs, a faulty or compromised kernel instance could still
be able to threaten the platform correctness.
\note{
Conventional OS design, supported by enforcement mechanisms accepting
changes from any single kernel instance, equates control over
resources with the possibility of directly or indirectly obtaining
access to the former. 
}
For example, by manipulating page tables, any low-level OS kernel
instance can install virtual-to-physical address mappings to any
resource in the platform's memory map and access it through this
mapping. 
\note{
Of course, it is imaginable to confine kernel access to a
boot-time fixed partition, in which case the adversary has to access
the resource indirectly through a compromised application whose
virtual-to-physical mappings the kernel controls.
}
Of course, a trusted underlying layer could solve this issue (e.g., by
mediating page-table access). However, whether this layer is software,
as in the Inktag kernel~\cite{Hofmann:2013:ISA:2499368.2451146}) or
firmware, as in Intel SGX~\cite{intel_sgx}), it becomes a single point of
failure for the platform.

\note{
In this work, we propose a paradigm shift in the latitude with which
low-level software, when faulty, can interact with the platform's
privilege mechanisms.
So far, a single instance of such software is always able
to configure privilege mechanisms, even when compromised.  
}

\midir provides a further level of protection, whereby the designer
can constrain access to the platform reconfiguration, by allowing a
particular mechanism, its registers and data structures to be only
effected in a consensual manner, through a voter.
As with general voting, discussed in
Sec.~\ref{sec:voted_invocation}, these voted accesses will normally
correspond to the implementation of detection or tolerance strategies,
in this case, directed to the protection against threats on the
platform itself.
In Fig.\ref{fig:invocation}(b), in green colour, we represent such a
flow of reconfiguration of a platform capability register in tile A's
\bluedot.
Exemplifying with {\em f-out-of-n} error masking in a replicated
low-level kernel,
%
several replicas make the reconfiguration request, which is voted
(green voter). The result from the voter is wired through a special \bluedot
capability configuration interface to the concerned capability
register, masking the presence of up to $f$ faulty replicas.

\note{
Provided no more than $f$ of these replicas become simultaneously
compromised, faulty replicas can no longer obtain access to platform
resources unless this access has been legitimated by at least one
correct replica.  Faults are contained to the resources accessible by
such a compromised replica.  

Note that the mechanism is not mandatory: disabling voting
(e.g. setting the fault threshold $f$ to zero) retains legacy
compatibility.
}

\section{Towards Fault and Intrusion Tolerant \\Microhypervisors}
\label{sec:hypervisor}

We now turn our attention to the construction of \midir-aware FIT
microhypervisors, such as suggested in Fig.~\ref{fig:architecture}. 
Hypervisor replicas execute on dedicated
tiles, from where they remotely configure the privileges of applications
executing on other tiles. Most of the other common OS-functionality
(e.g., context switching, inter-process communication, (non-critical)
device access, etc.) can be left to the application and its
kernel-support libraries.

\midir gives the designer latitude to use incremental levels of
protection for individual operations or sets thereof.  On one
extreme, configurations may be allowed where all accesses are direct,
and thus unprotected by voting.
%


%
On the other extreme, the highest level of protection, while retaining
the flexibility of a manycore system, eliminates all software-level
single points of failure\footnote{Modulo \midir's \bluedot, which,
  justified through its simplicity, we assume will not fail.} by
subjecting all critical operations to voting. We focus on this facet.
The replicated microhypervisor offers a system-call interface executed
by its replicas, entering a service loop and maintaining data
structures used to handle system call requests, which they receive
from applications, other replicas (e.g., requesting a privilege they
lack for executing a system call) or from hardware (e.g., triggered by
device interrupts).

Remembering that the unit of fault containment in \midir is the tile
(equivalent to a node in a distributed system) the essential
requirement for a fault tolerant microhypervisor design is that the
replicas behind critical operations are placed in different tiles,
such that they communicate by messages, are subject to \bluedot access
control, and converge on the necessary
votes as dictated by the algorithm.
In order to fully enjoy the baseline functionality provided by \midir,
a few additional design principles should be followed:
\begin{itemize}
  \item {\bf P.1 \emph{Impersonation prevention:}} Correct replicas
    must deny any operation with a replica identifier that is already
    in use (\bluedot voting relies on identifying the individual
    replicas through their capability).

  \item {\bf P.2 \emph{Bypass prevention}} Correct
    replicas must deny any operation attempting to grant direct write
    access to a consensual-update-only object
    (Sec.\ref{sec:privilege_reversion}).
 \end{itemize}

\note{
  \item {\bf P.3 \emph{Protecting replica integrity:}} Fault
    independence is obviously violated if faulty replicas can modify
    correct replicas, their code or critical data
    structures. Therefore, no correct replica must agree to the
    installation of capabilities that convey direct write access to
    these resources. Moreover, they must not agree to modifying these
    objects consensually other than to recover the respective replica.
}

\note{
Naturally, we cannot solely rely on hardware to retain this
flexibility, in particular if over time the same resources (e.g.,
memory or a tile) should be used for non-critical and critical
components (including, e.g., hypervisor replicas evading from
permanent damage). Instead, replicas must consensually maintain the
following system-level properties, to prevent replicas from obtaining
the possibility to cause damage once they become faulty.
}

\note{
On the other extreme, the highest level of protection, eliminating all
software-level single points of failure, implies that the system is
configured such that the following invariants are preserved initially
and during execution:

\begin{itemize}
\item {\bf Inv.1 \emph{Impersonation prevention:}} no correct
  hypervisor replica agrees to installing in different tiles,
  capabilities to the same voter with the same replica identifier.
\item {\bf Inv.2 \emph{Bypass prevention:}} no correct replica agrees
  to installing capabilities to consensually updated resources that
  grant direct write access\footnote{Doing so would bypass
    voting. \bluedot's reconfiguration interface cannot directly be
    addressed by a capability; accesses must pass through a voter.}.
\item {\bf Inv.3 \emph{Replica protection:}} no correct replica agrees
  to installing capabilities that could be used to damage other
  replicas. This includes among others capabilities authorizing writes
  to a replica's code and data segment.
\end{itemize}

With these invariants established, the microhypervisor (as usual)
offers a system-call interface executed by its replicas, entering a
service loop and maintaining data structures used to handle system
call requests, which they receive from applications, other replicas
(e.g., requesting a privilege they lack for executing a system call)
or from hardware (e.g., triggered by device interrupts).
}


Let us illustrate the design with the example of reallocating the tile
to a different application.
Signaling the tile, an application-specific library
may save the state necessary to resume execution (e.g., utilizing
memory assigned for this purpose). The actual switch then proceeds by
resetting the tile
followed by installing the capabilities the new application's library
needs, in order to load its state.  Obviously, reset (and, as we have seen,
privilege change) is a critical operation, which must be performed
consensually to prevent compromised kernel replicas from prematurely
stopping applications.
Channeling such critical operations to voters and confining
access with capabilities prevents faulty replicas from causing harm,
since, as long as no more than $f$ replicas become compromised, a
correct majority out of the $n = 2f + 1$ replicas will outvote these operations. This turns system-call
execution into updates of replicated state and a sequence of voted
operations, which we shall later call \emph{subordinate votes}. This
works as well with any other replicated critical software, even
firmware such as in SGX (e.g., preventing enclave misconfiguration) or
device drivers, when interacting with the physical world.
Replies to system calls must also be voted upon, given that hypervisor
replicas, by nature, act on behalf of multiple applications, possibly
storing information of one that must not be revealed to others.

\note{
In a direct response (like in
MinBFT~\cite{minbft}),
compromised replicas could leak this confidential
information\footnote{
  Notice, that we do not yet address the timing channels involved when
  encoding information into the time when a replica contributes its
  vote.}.
}

The above is of course true provided replicas have reached agreement on the
system call to execute and on the parameters with which the client
application has invoked this call. A further role of the service loop
is therefore to reach consensus on system call execution order and parameters.
From our evaluation (Sec.~\ref{sec:evaluation}) we found that \midir's
support for consensually executing critical operations
also provides for accelerating
the BFT protocol that the kernel replicas must execute to reach this
agreement.

\subsection{Consensual System Calls}
\label{sec:system_calls}

\begin{figure}
\begin{center}
\includegraphics[width=0.85\columnwidth]{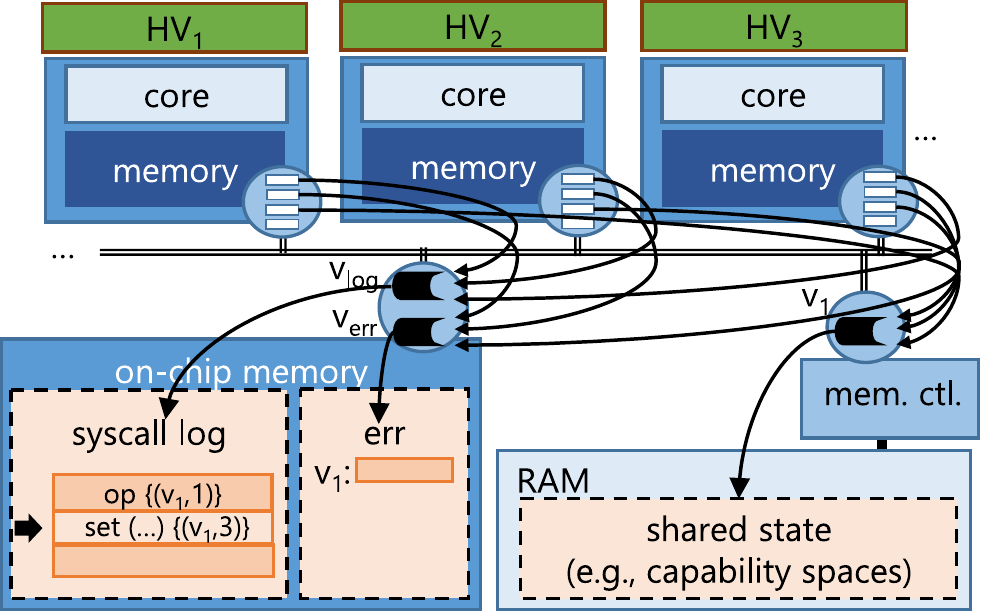}
\caption{Read-shared, consensually updated data structures used by the
  kernel: system calls are recorded in the syscall log, the error log
  keeps voting error information, a capability space holds an application's
  capabilities (Sec.\ref{sec:evaluation}).}
\label{fig:kernel_memory}
\end{center}
\vspace{-.5cm}
\end{figure}

Fig.\ref{fig:kernel_memory} provides a more detailed picture of how
\bluedot's voters and capability registers contribute to a FIT
hypervisor's service loop reaching consensus on the system call to
execute.


The service loop utilizes two data structures:
a consensually updated ringbuffer ---
the \emph{syscall log} --- records agreed upon system calls and its
parameters to give kernel replicas the opportunity to learn about those 
agreed upon.
%
Otherwise, this information would only be
available to the agreeing quorum of $f+1$ replicas and if faulty
replicas participate there, but refuse to execute the system call
later on, too few correct replicas would have obtained this knowledge
to complete the system call.
Similarly, the service loop utilizes an \emph{error log} to protect
error information from getting lost in premature resets of the voter.
\note{
We dimension the syscall log large enough to not wrap around before
all replicas have been rejuvenated proactively. Alternatively, log
entries can be garbage-collected after checkpoints. 
}
Updates of the syscall and error logs are made through dedicated
voters: $v_{\mathit{log}}$ and $v_{\mathit{err}}$, respectively.

Macroscopically, clients place system-call requests in authentic
buffers, which the kernel replicas poll\footnote
{
Sleep/wake protocols can be used in periods where no requests are pending.
}
for new requests. Consensual privilege change allows creating such buffers 
by granting write access to a single client, but to no kernel replica. The 
leading kernel replica proposes one such system call by initiating a
vote with $v_{\mathit{log}}$, which followers introspect and agree or
deny. Once written to the syscall log, replicas proceed by executing
the system call and the votes for its critical operations. We call
these \emph{subordinate votes} as they depend on the main vote,
logging the system call. That is, no correct replica will engage in a
subordinate vote unless the system call has been logged. Subordinate
votes include at least replying to the client and advancing the
syscall log to the next free slot. They are performed utilizing 
a set of
voters $V = \{v_1, \ldots\}$ that is disjoint from
$\{v_{\mathit{log}}, v_{\mathit{err}}\}$.

We make no assumptions on the order in which replicas update their
local state (even transactional or speculative updates are
imaginable). However, to simplify tracing the progress of the system
call (and in turn the code that late or rebooted
replicas have to execute to catch up), we require subordinate votes to
be executed in the same order by all replicas and assume that this
order is completely specified by the system-call parameters.

Our rationale for agreeing on the system call first is to
circumvent a fundamental problem of consensus protocols without
authenticators: the impossibility to diagnose faults if messages can
be altered during multicast
operations~\cite{Lamport+Shostak+Pease:1982}. In our setting,
cryptographic operations would come at overproportionally high costs
relative to the speed of the transport medium (the NoC).
We therefore avoid sending unforgeable
authentication tokens (e.g., HMACs)
and instead
exploit the authentication we obtain from a client being the single
writer of its request buffer.
Additionally, clients maintain write access to their request
buffers. Thus, they can change the request after the leader has
proposed it, but before followers validate it, which makes it
impossible for followers to distinguish whether the leader proposed a
wrong system call or whether the leader proposed the client's original
suggestion, but the client changed it afterwards. In consequence, they
cannot differentiate faulty clients from faulty leaders to provably
identify the leader as faulty.
We omit error diagnosis for the system-call vote to regain it when we
need it: in the subordinate votes for reaching agreement on critical
operations.


The following details the protocols the hypervisor replicas execute to
reach consensus on and execute system calls. Leveraging the generic
voting pattern in Fig.\ref{fig:generic_pattern}, replicas
first reach agreement on the system call
(Fig.\ref{fig:consensus_protocol_syscall}) to then
consensually perform critical updates during its execution
(Fig.\ref{fig:subordinate}).

\subsection{Generic Voting Pattern}
\label{sec:general_voting}

\begin{figure}
\begin{lstlisting}
1  agreement:
2    $\mathit{seq}_i := v_i.\mathit{seq}$
3    if (replica $= \mathit{seq}_i \mathit{mod}\, n$) {
4      // leader
5      $v_i$.propose($op$, $\mathit{seq}_i$)
6    } else {
7      // follower
8      wait for leader proposal: $op$
9      validate op
10     if (valid) $v_i$.confirm($op$, $\mathit{seq}_i$)
11       else     $v_i$.decline($op$, $\mathit{seq}_i$)
12   }
13   // all
14   wait for $f+1$ replicas to
15       agree/disagree/timeout
\end{lstlisting}
\vspace{-2mm}
\caption{Generic voting pattern used in the service loop and when executing system calls.}
\label{fig:generic_pattern}
\vspace{-2mm}
\end{figure}

Fig.\ref{fig:generic_pattern} shows the generic pattern and how
replicas interact with voters. Evaluating the sequence number
$v_i.\mathit{seq}$ of voter $v_i$, replicas identify the leader as the
replica with identifier $v_i.\mathit{seq}\, \mathit{mod}\, n$ in its
capability. The leader proposes a request by invoking its vote capability
to write operation $op$ to its voter buffer, which the voter prevents
from being changed once the leader marks this proposal as
complete. Followers wait for the leader to complete its proposal to
then validate the operation and express their agreement/disagreement
(by submitting the operation they saw or by writing
the corresponding value to the agreement vector (see Sec.\ref{sec:implementation})).

\subsection{System Call Vote}
\label{sec:syscalls}

\begin{figure}
\begin{center}
\begin{lstlisting}
16  client $c_k$:
17    write $m :=$ syscall opcode + parameters
18        to $c_k$'s request buffer
19    wait for reply in $c_k$'s response buffer@\\[1mm]@
20  hypervisor replica $HV_i$:
21    service loop:
22        poll all client buffers
23        remember new request $(m, c_k)$ as pending@\\[1mm]@
24      on pending request:
25        // leader
26        $(m, c_k) :=$ pending.remove_head
27        if ($m$ is invalid syscall)
28          skip to next pending request
29        $\mathit{VS} := \emptyset$
30        for each voter $v_i$ used to execute $m$
31          // collect voter sequence numbers
32          introspect $v_i$ to read $\mathit{seq}_i := v_i.\mathit{seq}$
33          $\mathit{VS} := \mathit{VS} \cup \{(v_i, \mathit{seq}_i)\}$
34        // follower
35        if (pending requests $\neq \emptyset$)  
36            set timeout
37        // all
38        $v_{\mathit{log}}$.agree_on (``write(log, $\langle m, c_k, \mathit{VS}\rangle$)'')
39           with validate $:=$
40             ($\mathit{m} \neq$ request from client $c_k$) ||
41             ($v_{\mathit{log}}.\mathit{seq} \neq \mathit{seq}_{\mathit{log}}$) ||
42             ($\mathit{seq}_v \neq v.\mathit{seq}$, where $(v, \mathit{seq}_v) \in \mathit{VS}$))
43        if (at least one replica disagrees)
44          $v_{\mathit{log}}$.vote_for_reset()
45        if (not $f+1$ agreement)
46          repeat vote
47        execute $m$
\end{lstlisting}
\vspace{-2mm}
\caption{Service loop - Phase 1: agree on next system call to execute}
\label{fig:consensus_protocol_syscall}
\end{center}
\vspace{-5mm}
\end{figure}

In Phase~1, replicas first agree on the
system call to execute following the generic pattern above. 
In Phase~2, they then vote on critical operations.
Fig.\ref{fig:consensus_protocol_syscall} shows the pseudocode for
system-call agreement. Lines~16--23 illustrate the client invocation
pattern discussed above. The leader
selects a pending system call (Line~26) with
valid opcode (Line~27) and prepares the entry to log. To prevent
equivocation during subordinate votes (e.g., attempts to trick a
replica into proposing the next system call without completing the
current one), we enforce some additional principles:
\begin{itemize}
\item {\bf P.3 \emph{Coordinated subordinate votes:}} correct
  replicas vote only on subordinate voters ($v_i \in V$) to execute
  the current system call.
%
\item {\bf P.4 \emph{Presence of correct replica:}} no voted
  operation succeeds without at least one correct replica.
\end{itemize}
We enforce P.4 
by requiring
quorums of at least $f+1$ matching votes, while preventing
impersonation (c.f., P.1 in Sec.\ref{sec:hypervisor}).
In combination, these  principles ensure that subordinate voters $v_i
\in V$ will keep their state while in Phase~1 (including their
sequence numbers). By agreeing, alongside the system call, on the
first sequence number of all voters used in this system call
(collected in Lines~29--33 in the set $\mathit{VS}$ and validated in
Line~42), we ensure that all replicas know all sequence numbers to
start with in subordinate votes, even if they have been lagging
behind. In the absence of errors, the $j^{th}$ subordinate vote on
$v_i$ will be executed with sequence number $\mathit{seq}_i + j$,
assuming $(v_i, \mathit{seq}_i) \in VS$ was the start sequence number
of $v_i$. This agreement on the initial sequence number then allows
for a simpler progress tracking in Phase~2, when executing subordinate
votes.

Because of the impossibility in Sec.\ref{sec:system_calls},
system-call votes operate with reduced error diagnostics:
replicas reset $v_{\mathit{log}}$ if it got suspended after
disagreement (Lines~43, 44) and repeat votes for pending system calls
unless they fail for all client-leader combinations, in which case they 
exclude this client.

\subsection{Subordinate Votes}
\label{sec:subordinate_votes}

\begin{figure}
\begin{center}
\begin{lstlisting}
48  $HV_i$.vote (log, $v_i$, $\mathit{seq}_i$, req, $m$, dest) {
49    if (syscall_log.log $\neq$ log) 
50      return success
51    if ($v_i.\mathit{seq} \neq \mathit{seq}_i$)
52      if ((err[$v_i$].log $\neq$ log) ||
53          (err[$v_i$].req $\neq$ req) ||
54          (err[$v_i$].$\mathit{eseq} > \mathit{seq}_i + 1$))
55        return success
56      push_error_and_reset_voter
57      if (!err[$v_i$].success)
58        repeat vote with $\mathit{seq}_i + 1$
59    // @{\color{red}$HV_i$}@ is up to speed with the others
60    $v_i$.agree_on(``write(dest, $m$)'') with $\mathit{seq}_i$
61         and validate $:=$ ($m$, dest) is valid
62    if (at least one replica disagrees)
63      push_error_and_reset_voter
64      initiate recovery
65    if ($f+1$ agreement)
66      return success
67    repeat vote with $\mathit{seq}_i + 1$
68  } @\\[.1mm]@
69  push_error_and_reset_voter:
70    error $:=$ introspect($v_i$)
71    $v_{\mathit{err}}$.agree_on(``write(err[$v_i$], error)'')
72        with validate $:=$ 
73          adjust own error information
74          (proposed error $=$ own error)
75    if (error vote fails)
76      $v_{\mathit{err}}$.vote_for_reset($\mathit{eseq}$)
77      repeat pushing the error
78    $v_i$.vote_for_reset($\mathit{seq}_i$)
\end{lstlisting}
\vspace{-2mm}
\caption{System call execution - Phase 2: subordinate votes and error handling}
\label{fig:subordinate}
\end{center}
\vspace{-5mm}
\end{figure}

The code for executing subordinate votes in Fig.\ref{fig:subordinate}
has to solve two problems:
(i) preserve determinism despite errors and
(ii) prevent replicas from prematurely resetting voters.
%
From reaching agreement on the system call, we know that the first
subordinate vote on $v_i$ starts with $\mathit{seq}_i$ because $(v_i,
\mathit{seq}_i) \in \mathit{VS}$. As such, without errors, the $j^{th}$
subordinate vote on $v_i$ happens with sequence number
$\mathit{seq}_i + j$. The same applies to votes with at least one
disagreeing replica that all received $f+1$ agreement because, after
the voter resets (Line~62), they are not repeated (Line~66). The key for
lagging replicas to catch up in case of error is to make sure they
learn about all errors, so that they know how many times a vote was
repeated and when it was successful. Assume the $k^{th}$ subordinate
vote ($k < j$) was the last to fail with $\mathit{seq}_i^k$, then $k$
completed with $\mathit{seq}_i^k+1$ and the system call progressed to
subordinate request $j$ if $v_i.\mathit{seq} - \mathit{seq}_i^k = j -
k$.

Solutions to the second problem address the point that all replicas
must learn about errors. With $n = 2f+1$ and $|Q| = f+1$, up to $n - |Q| = f$
replicas may lag behind while the remaining $|Q|$ progressed to
another subordinate request or even to another system call. In
particular, faulty replicas may fail a subordinate vote but
agree to reset the voter, which erases the error information about the
failed vote from the voter and leaves behind as few as a single
correct replica to know about the error.
%
This scenario occurs if $f$ faulty and one correct replica resets the
voter before others diagnosed it.
%
Clearly, without costly cryptographic information, the honest replica
cannot convince
others 
about what has happened.  The following design principle solves
this problem by preventing premature resets before error information
is pushed to the error log.
\begin{itemize}
\item {\bf P.5 \emph{No reset before error logging:}} correct
  replicas reset subordinate voters only after the error got 
  logged.
\end{itemize}
This error state contains
information about the current system call, i.e.: the system-call entry
$\mathit{log}$; the subordinate vote $\mathit{req}$; the sequence
number of the voter $v_i$; the point where it failed $\mathit{eseq}$ and
which replicas agreed/disagreed. In consequence, lagging replicas can
validate if the current subordinate vote succeeded (Lines~52--55) and,
if not, who was responsible for it to fail.
Voter $v_i$ prevents destructive writes until it is reset, which P.5
and P.4 ensure happens only after error information was written to
the log. Non-destructive writes are updates of empty
buffers respectively updates of the agreement vector from timeout to
agree/disagree and from empty to any of these three.

The argument for why the problem does not recur with the nested vote for
logging the error state is as follows:
(i) The state to push is held in the voter $v_i$. Therefore, even if a
replica lags behind, finding $v_i$ suspended, it knows what
information to write to the log. (ii) Because of P.5, and because at
least $f+1$ replicas are required (P.4) for votes to succeed, the only way to make
progress is by writing correct error information. Therefore, either
faulty replicas agree to writing correct error information or
eventually correct replicas catch up and write correct
information. The exact information seen by the replicas may differ
depending on the time they read it, i.e., in late reads, more replicas may have expressed their consent or disagreement. However, it will always contain at
least the consensual result of the vote (i.e., whether $f+1$ replicas
agree, disagree or timed out) and, in the former two cases, it
identifies at least one replica that diverges from the majority (the
leader, in case of $f+1$ disagreement). This replica is proven faulty.
Followers, reading error information after the leader and finding
proposals of additional replicas, downgrade their own information to
that of the leader after validating it as described above (Line~73).
%
Repeating the vote while rotating the leader ensures that valid error information
is proposed latest after $f$ retries. It then suffices to reset
$v_{\mathit{err}}$, whenever it becomes suspended (Line~76). Once
error information is pushed, replicas vote to reset the voter $v_i$
for the subordinate vote (Line~78) and continue executing it.
%

\section{Implementation}
\label{sec:implementation}

The implementation of capability invocation is standard
(c.f.\cite{needham:cap}): \bluedot intercepts external operations,
looks up the capability in the capability register file, and
forwards the operation to the NoC after the privilege check succeeds,
silently dropping the operation otherwise. Replica IDs are
communicated as labels in the capability~\cite{Hardy:1985:KA:858336.858337}, which \bluedot inserts as
additional parameter into the operation.

Our
voter implementation is driven by the following
considerations and their impact on functional simplicity.

\subsection{Buffered vs. Unbuffered Votes}
Perhaps most impactful is the decision to buffer votes 
to allow replicas to make their proposals without first
having to synchronize on the time when the signal for such a vote must
be held. Although buffering increases the complexity of the voter,
it decouples replicas, allowing them to act in a partially synchronous
fashion and, as long as different voters are used, even partially
out-of-order\footnote{
  To simplify monitoring of the progress of a system call, we shall later
  require that all replicas execute the critical operations of each
  system call in the same order. Operations of different system calls
  need not be constrained in this way, and, at the cost of a
  more complex progress tracking, this requirement can be further
  relaxed to: same order as far as a single voter is concerned.}.
Buffering votes is ideal in a NoC architecture, since votes are
transmitted as normal messages. Tiles can continue executing once
the message is sent.
We therefore implement voters to contain buffers for storing proposals
from the different replicas for the current vote executed with this
voter.

\begin{figure}
	\begin{center}
		\includegraphics[width=.75\columnwidth]{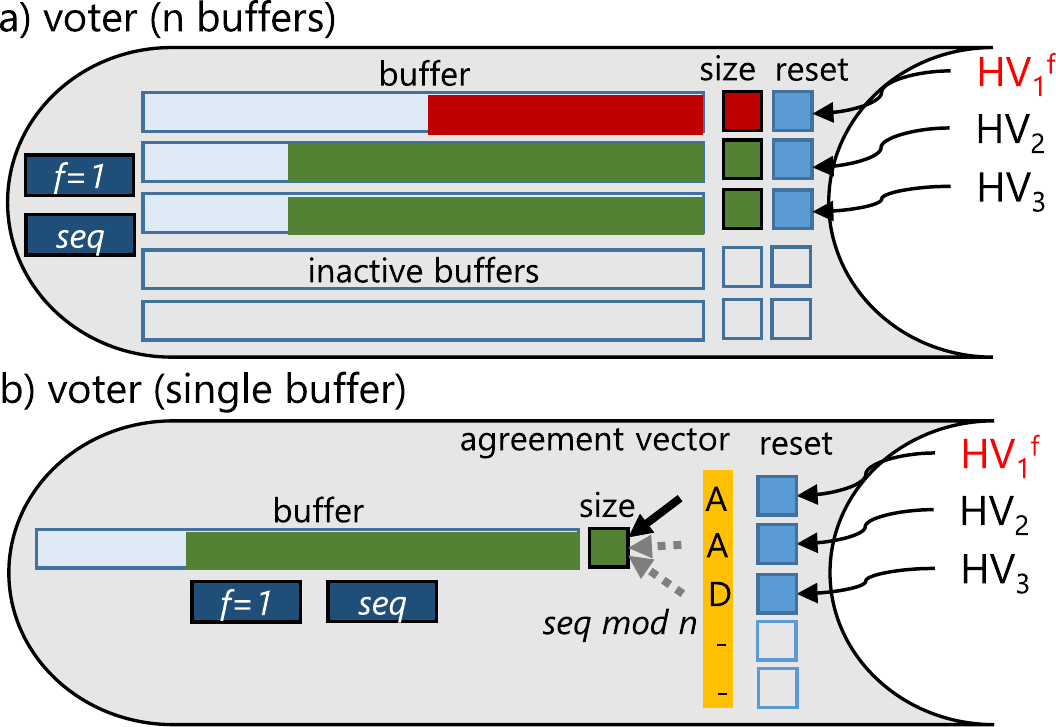}
		\caption{Internal structure of a voter. One, resp. $n$ buffers
			hold the message of replicas to vote upon and $\mathit{size}$
			its length. $f$ defines the fault threshold, $\mathit{seq}$ is a
			voter maintained sequence number. The 
			agreement and reset vector are described below.}
		\label{fig:voter}
	\end{center}
	\vspace{-.5cm}
\end{figure}

\subsection{Immediate vs. Deferred Masking}
A similarly impactful decision is whether voters should be able to
mask faults immediately. Alternatively, voting can be repeated until a
valid proposal is made. The consequences, besides time to agreement,
are the amount of memory needed for buffering votes vs.\ the
complexity of the voter logic.

To mask faults and reach agreement immediately after $|Q| = f + 1$
matching proposals arrive, the voter needs to buffer suggestions from
at least $f+1$ replicas.
%
%
Since up to $f$ such messages may be wrong and because the voter can
only find out after receiving $f+1$ matches, buffer space for at least
$f+1$ messages is needed to not have to repeat the vote.

We implemented two variants of \bluedot voters to evaluate the
resource/performance trade-off at the two extremes of this spectrum.
Our $n$-buffer variant (Fig.\ref{fig:voter} a) implements one
message buffer per replica. Each time a message arrives, it is
compared against all other stored messages and the operation applied
once $f+1$ buffers match.
%
Our single-buffer variant (Fig.\ref{fig:voter} b) trades agreement
time for a more resource-efficient implementation: there is only one
buffer; and only the current leader is granted write access to this
buffer.
The single-buffer voter follows a leader-follower voting scheme, with
the leader proposing a vote and followers validating this proposal. To
prevent inconsistency, the voter prevents modification of the leader
proposal once the leader marks the proposal as ready. This allows
follower replicas to introspect the stored message and express their
agreement/disagreement. For this purpose, the single-buffer voter
implements an agreement vector with one (initially empty: $-$)
tri-state cell for each replica to express agreement $A$ or
disagreement $D$.
%
Now, one of three things may happen when replicas propose:
\begin{enumerate}
\item [(i)] a majority of $f+1$ or more replicas disagree with the
  leader proposal. In this case, the leader proposal is considered
  invalid and the operation is not applied; or 
\item [(ii)] a majority of at least $f+1$ replicas agree. In this
  case, the proposal is accepted and the voter applies the operation
  in its buffer.
\item [(iii)] the operation times out without a majority of replicas
  agreeing / disagreeing. In this case, the replicas record this error
  and repeat the vote after rotating to the next leader.
\end{enumerate}

The $n$-buffer version requires logic circuits for pairwise buffer
comparison whereas in the single-buffer version a 2 data-bit majority
gate over the agreement vector suffices.

\subsection{Internal vs. External Error Handling}
\label{sec:voter_error}

The third question is whether the voter itself should include
provisions for diagnosing errors and for informing replicas about
them. Errors are detected when one replica diverges with the majority
decision.
Voter-initiated error handling translates to the voter tracing
back to the voting replicas' cores to identify where to deliver
error-handling interrupts. The expected complexity discourages such a
solution. We therefore offload error handling to software
and support replicas 
by a means to track progress (the sequence
number $\mathit{seq}$) and by suspending voting after detecting a
mismatch.
In this situation, $\mathit{seq}$ does not advance but the voter may still apply the operation
(in case of $f+1$ agreement). Replicas
introspect
the voter
registers and buffers to diagnose the error, by looking for divergences.

To resume execution of suspended voters, replicas reset the voter,
which clears all buffers and the agreement and reset vectors and
advances the sequence number by one. Reset itself is a voted operation
over the reset vector, which contains one bit per replica. The voter
resets once $f+1$ bits in this vector are set.
Although this quorum guarantees that at least one correct replica
agrees to resetting the voter, it does not prevent faulty replicas
from resetting the voter prematurely, that is, before all correct
replicas were able to retrieve the error state. P.5 and the
protocol in Sec.\ref{sec:subordinate_votes} handles this corner case.

\subsection{Dimensioning Voters}
\label{sec:voter_dimension}

The last question we discuss here is: for how many faults should the
voter hardware be laid out. Since we aim at implementing voters in
silicon,
we have to make this choice at system design time to dimension buffers
and vectors large enough for the maximum number of faults to
tolerate ($f_{\mathit{max}}$). However, to not always have to execute at this maximum
replication degree, a fault threshold $f \le f_{\mathit{max}}$ of voters can be
configured at boot time.
For instance, 
if the system should tolerate up
to $f_{\mathit{max}} = 3$ faults, it needs to be dimensioned to have
$n_{\mathit{max}} = 2 f_{\mathit{max}} + 1 = 7$ fields in the vectors
(and $n_{\mathit{max}}$ buffers, assuming $n$-buffer voters). This voter can
be operated at any fault threshold $0 \le f \le f_{\mathit{max}}$.

The voter design has been kept simple enough, and decoupled enough
from the surrounding logic. As such, we can expect with high
confidence that \bluedot can be implemented and shown correct, as well
as stay functional even when the tile it is associated with fails.
%

\section{Evaluation}
\label{sec:evaluation}

As an early validation of our proposal, we have implemented \bluedot
in both voter variants in VHDL on a Zynq-7 ZC702 Evaluation Board.
We instantiated 3 Microblaze cores as tiles, running at 50~MHz, each with
one \bluedot, connecting the tiles through \bluedot with an
AXI interconnect (serving as the NoC). We measured the performance of the service loop
(Fig.~\ref{fig:consensus_protocol_syscall}) to agree on and execute client-invoked
system calls for granting and priming capabilities.  Grant ({\tt
  L4.map}~\cite{Liedtke:sosp:l4:1995}) copies capabilities between
capability spaces and prepares for later revocation. 
Prime consensually copies a capability from the client's capability space
into a \bluedot capability register, where it is ready for invocation.
We have measured the performance of grant and prime in two different
implementations of capability spaces\footnote{Container object for an application's capabilities.}: 
(i) as a private data
structure in each replica, requiring, in the case of prime, only the vote to
install capabilities
and two further to reply to the client and mark the system call as
finished; and (ii) as a read-shared, consensually updated data
structure, trading off speed for a smaller memory footprint by
introducing additional votes for track keeping.

As baselines, we
compare to a cross-tile invoked singleton kernel (horizontal line),
executing the same system calls on its private state, with 1637~cycles
for \emph{grant} (1977~cycles for \emph{prime}) and to a shared-memory
variant of MinBFT\footnote{
  We omit client signatures in favor of
  authentic buffers, but implement UIs with HMACs. USIGs can be
  accessed without overhead.}
requiring 242824~cycles to agree on a system call.
Our agreement protocol outperforms MinBFT by one order of magnitude.


\subsubsection{Per-Replica Capability Space}
\label{sec:per_replica_cspace}

\begin{figure}
\centering
  \includegraphics[width=.8\columnwidth]{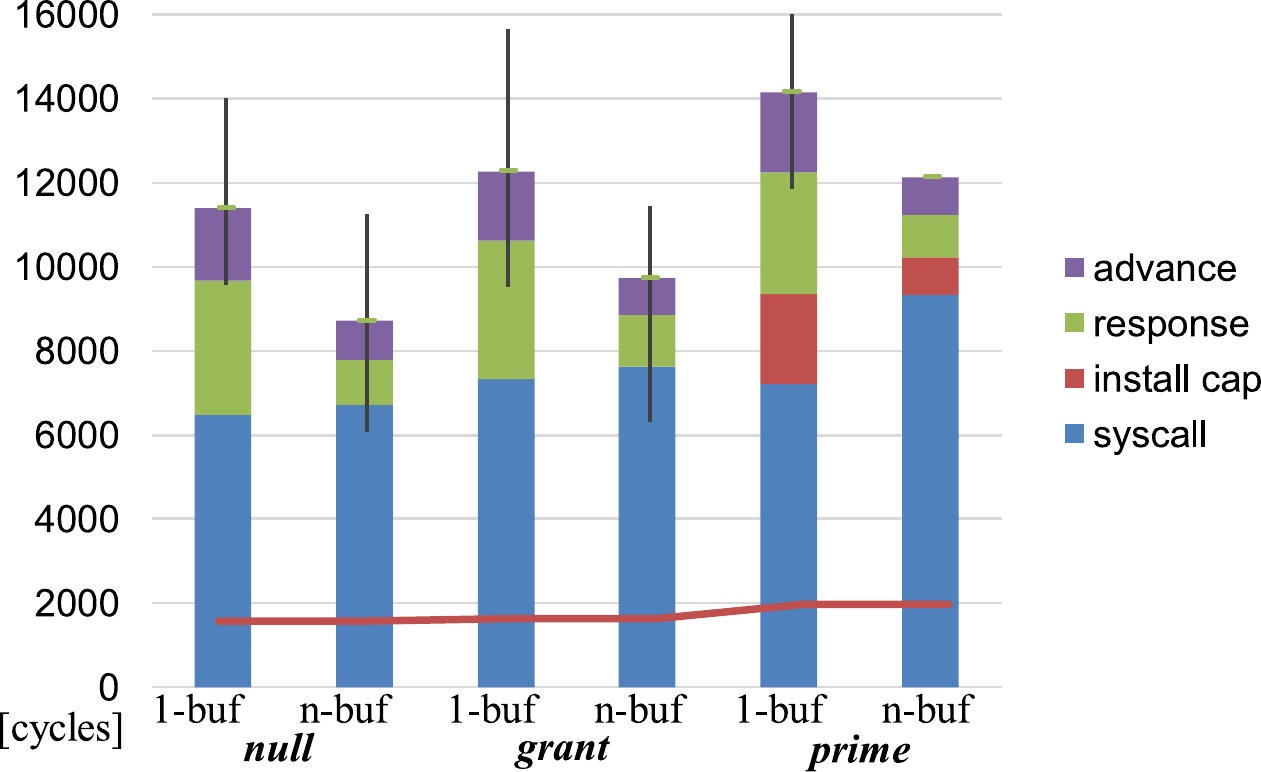}
  \caption{Average execution times of the three consensual
    system calls --- \emph{\bf null}, \emph{\bf grant} and \emph{\bf
      prime} --- when executed on a per-replica capability space
    implementation. System calls are broken down into the individual
    votes for agreeing on the system call and for performing the
    critical updates required. Shown are also the Q5 / Q95 percentile
    and the average costs of executing the respective system calls on
    a singleton-kernel.}
  \label{fig:per_replica_cspace}
\end{figure}

Figure~\ref{fig:per_replica_cspace} shows the average performance of
the \emph{\bf grant} and \emph{\bf prime} system calls in a per-replica
capability space implementation relative to the two baselines:
\emph{\bf null} and a singleton kernel instance performing these
system calls in a non-consensual manner. Shown are the system calls
broken down into individual votes and the Q5 / Q95 percentiles of the
overall measurements.

The minimal costs for learning about a system-call request and
executing it are 1571, 1637 and 1977 cycles on average for null, grant
and prime, respectively, which is the baseline of the singleton
kernel. System calls for the single buffer version have a factor $8.9$
-- $9.6$ increase, which can be explained due to the voter not
benefiting from caching. Whereas the singleton kernel merely has to
copy one request from the memory where the client core places it,
missing
in all caches in the process, following replicas have to poll
the voter to wait for the leader to make a proposal and then confirm
(or reject) the proposal made. Each such voter access amounts to costs
equivalent to a cache miss.

As can be seen, reaching agreement on the subordinate votes is much
faster, which is due to the fact that replicas already align
themselves when reaching agreement on the system call to execute.

In the n-buffer version of the voter, higher costs occur during the
agreement on the system call, which is due to the writing of the
complete request to the voter, not just setting a bit in its agreement
vector. However, subordinate votes are much faster, since replicas no
longer wait for the leader to make a proposal. Instead, they just
propose what should be written as critical operation.

\subsubsection{Consensually Updated Capability Space}
\label{sec:consensual_cspace}

\begin{figure}
\centering
  \includegraphics[width=\columnwidth]{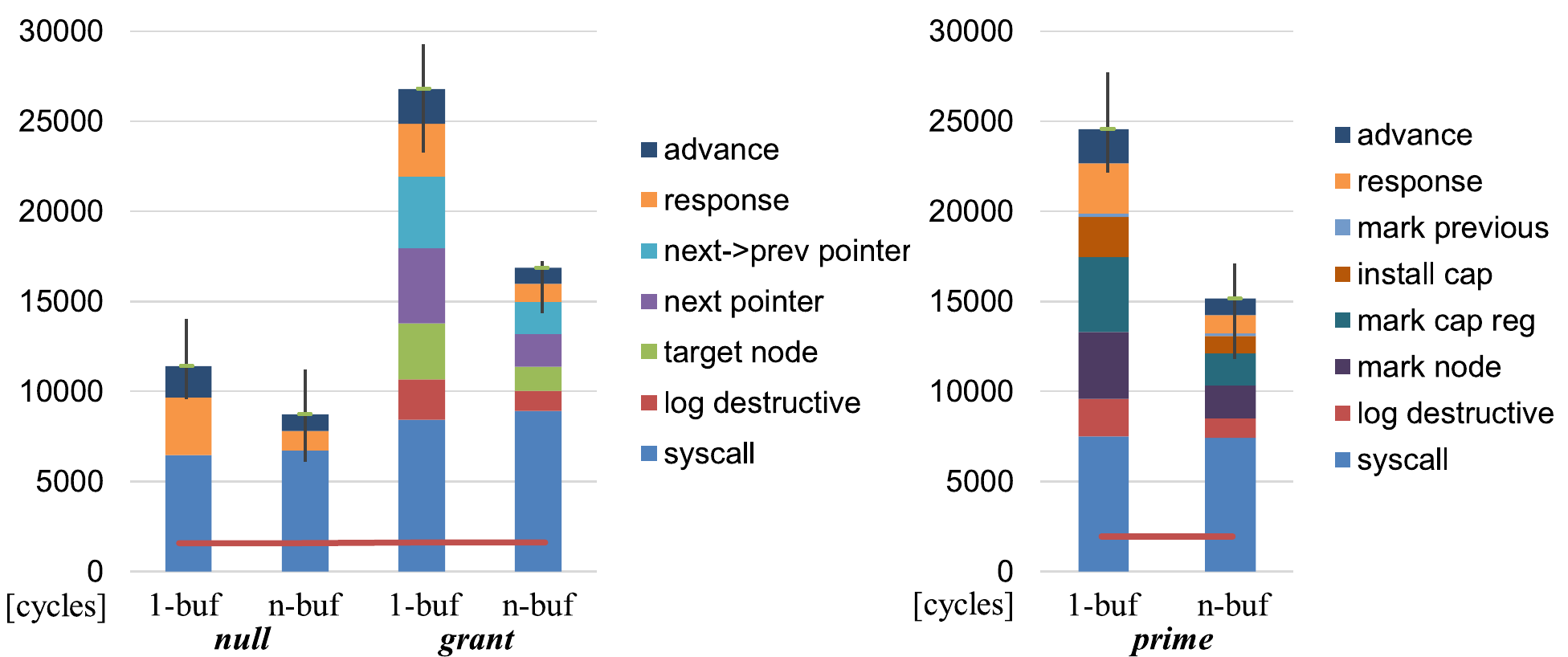}
  \caption{Average execution times of the three system calls for
    consensually updated capability spaces.}
  \label{fig:consensual_cspace}
\end{figure}

Figure~\ref{fig:consensual_cspace} shows a similar diagram as
Figure~\ref{fig:per_replica_cspace}, this time, however, for
consensually updated capability spaces. Granting and priming
capabilities now require additional votes to update the
data structure. 

Again, the 6.7 (/ 7.3) times slower performance relative to the singleton kernel can be explained due to the voter not benefiting from caching:

\emph{Singleton kernel:} System call execution is triggered by the
client writing to shared memory on one core and the kernel (on another
core) reading it.  From then on, all the operations happen locally in
the core of the kernel without any interaction with the
outside. Therefore, all memory operations aside from the invocation
and reply hit in the core's cache, which in our setting
responds within 1 cycle. The cross-core operations (invocation (1)
+ reply (2)) dominate these costs.

\emph{Replicated kernel:} System call execution starts as well with
invocation (1), but then, the leader needs to propose the request (2),
followers
validate it and (3) 
express
agreement (4) upon which the voter updates the memory and all replicas
wait for the vote to reach agreement (5). In (i), we then execute
locally, but for replying (to not introduce storage channels)
we have to repeat at least (4) + (5), assuming $n$-buffer voters.
As such, even without any delays, we have 7 cache misses vs. 2 in the singleton kernel
execution, hence a factor of 3.5. Additionally, more voter accesses are performed to read 
the sequence number, which we need for flow control.

\begin{figure}
\centering
{\footnotesize N-buffer voter}\\[1mm]
\includegraphics[height=3.5cm]{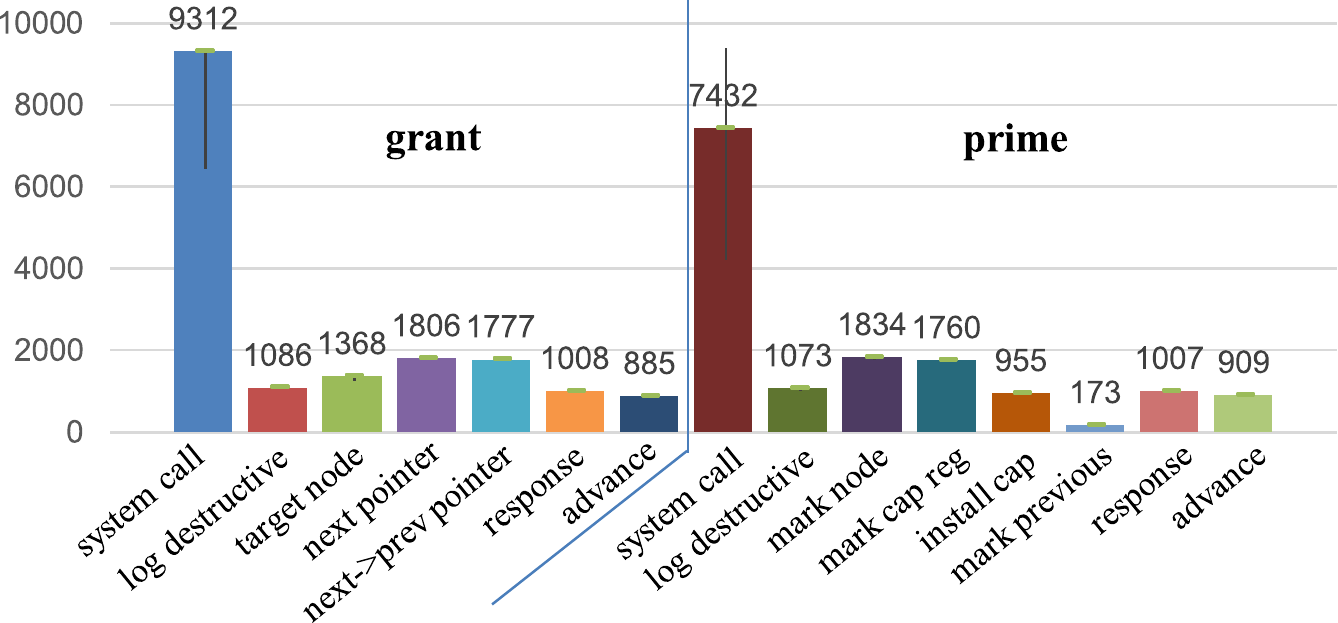}
  \caption{System calls broken down into individual votes. Shown are
    the Q5 and Q95 percentile for the main system call vote and each
    subordinate vote for n-buffer voters. The variations for
    single-buffer voters are similar.}
  \label{fig:variations}
\end{figure}

To confirm that variations in fact originate from the agreement on the
system call to execute, we have broken down system call execution into
their individual votes and measured their Q5 and Q95
percentile. Fig.~\ref{fig:variations} shows these values. As expected,
subordinate votes remain close to their average execution times,
whereas agreement on the system call varies significantly.

\begin{figure}
  \includegraphics[width=\columnwidth]{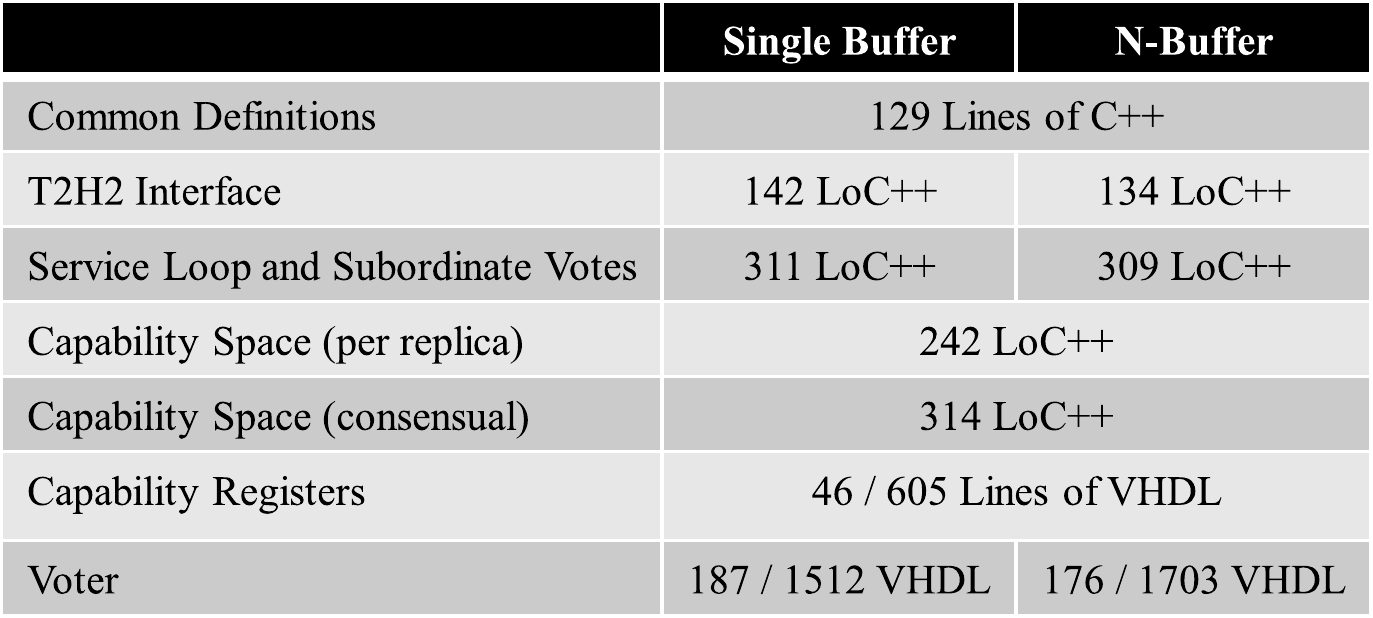}
  \caption{Code size in lines of C++ / VHDL code (logic/total).}
  \label{fig:code_size}
\end{figure}

Fig.~\ref{fig:code_size} lists the code size (excluding initialization)
for the service loop, for consensually executing critical operations
and for interfacing with the capability registers.
Also shown are the VHDL source lines of code for the logic and the
overall design of the voter and capability unit.
As can be seen, the amount of code that each replica executes for the
above grant and prime system call is well below 1000 lines of
code. Faults in this code are masked by the majority of replicas
outvoting faulty replicas in critical operations.
Similarly, the hardware overhead is just above 400 lines of VHDL code
for the logic plus 2411 lines of VHDL for connecting the logic to the
AXI interface and for mapping the corresponding internal signals.

\begin{figure}
\begin{center}
  \includegraphics[width=\columnwidth]{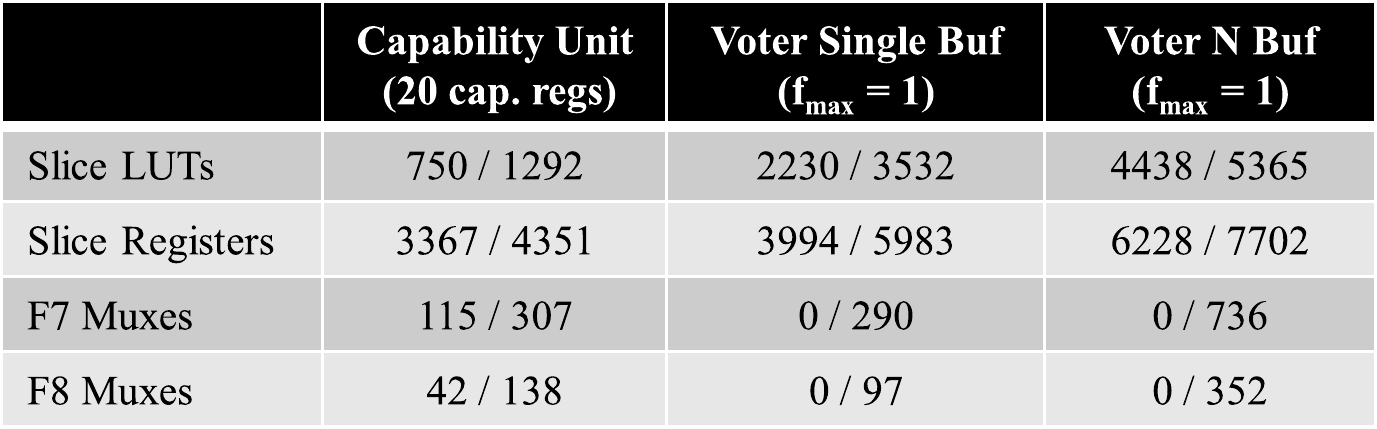}
  \caption{FPGA resources required by \bluedot (without / with AXI interface).}
  \label{fig:performance}
\end{center}
\end{figure}

Fig.\ref{fig:performance} shows the FPGA resources of the (post-synthesis)
implementation of our components.
LUTs are units with no state, used to implement the combinatorial logic; while
registers hold state, e.g, to keep buffer contents, but implement no logic.
Each F7 Mux (wide multiplexer) combines the outputs of two LUTs together, while F8 Muxes
combine the outputs of two F7 Muxes. 

Notice that the absolute resource requirement of
T2H2 will not increase significantly if more complex cores are to be controlled.
Hence, the relative overhead will shrink when more complex tiles are considered.

%
%
%
%
%

\section{Related Work}
\label{sec:related-work}

In this section, we present several classes of works that motivated
\midir: low-level approaches for detection and containment of
errors in low-level support
software; analyses of the evolution of defects in system support
software; 
attempts at preventing and/or
mitigating the resulting errors and potential failures; 
approaches to replication-based fault/intrusion tolerance and resilience.

Mitigation measures have been studied for detection and containment of
errors in OS and manycore support
software~\cite{5504713,Seshadri:2007:STH:1294261.1294294,romain}
through an underlying, assumed-trustworthy layer. However, they
still have a non-negligible complexity,
and in consequence, even a residual fault or vulnerability rate in these
supposedly trusted components may breach the platform's dependability
and security goal.  

In fact, as confirmed by~\cite{univis91357728}, ``simple'' components
with at least a few KLOCs have a non-negligible statistical fault
footprint.
%
Other studies~\cite{Ostrand:2002:DFL:566172.566181,Ostrand:2004:BUG:1007512.1007524} reveal between 1--16 bugs
per 1,000 lines of code go undetected before deployment, even in
well-tested software, and operating-system kernels form no
exception~\cite{ganapathi,Matias:2014:EES:2554850.2555021}.  Recent
insights~\cite{Palix:2014:FL:2642648.2619090} reveal that faults in
stateful core subsystems --- on which we focus here ---
outrank driver bugs in severity.
Minotaur introduces a toolkit to improve the analysis of software vulnerability
to hardware errors by leveraging concepts from software testing \cite{mahmoud2019minotaur}.

\note{
While early studies showed that most bugs concentrate in driver
code~\cite{chou}, more recent
insights~\cite{Palix:2014:FL:2642648.2619090} reveal that faults in
the stateful core
subsystems start to outrank driver bugs in severity. Faults of this
latter class can also occur in OS designs, facilitating
user-level drivers
(microkernels or -hypervisors),
on which we focus here.
%
}

%
Many approaches target operating
systems with the goal of improving their resilience against
faults. However, typically they protect either
applications~\cite{otherworld,Bolchini:2013:SFT:2467238.2467247,haft}
or specific OS
subsystems~\cite{Sundararaman:2010:MOS:1837915.1837919,Swift:2006:RDD:1189256.1189257,Zhou:2006:SSR:1267308.1267312,Shen_Elphinstone}
and only from accidental faults. Efforts for providing
whole-OS fault tolerance
include~\cite{Herder:2006:CHD:1170132.1170290,Nikolaev:2013:VOS:2517349.2522719,David:2008:CIR:1855741.1855746,Lenharth:2009:RDO:1508244.1508251,osiris,Govil:1999:CDR:319151.319162,david:os_attacks}.
Furthermore, the complexity of these recovery kernels is comparable to that of a
small hypervisor. For example, OSIRIS~\cite{osiris} directs OS
recovery to a 29 KLOC reliable computing base (RCB)~\cite{rcb},
roughly twice the size of
modern
microkernels~\cite{Liedtke:sosp:l4:1995,l4fiasco,Klein+al:sosp:sel4:2009,asmussen:m3}.
Again, this makes the likelihood of residual faults or vulnerabilities non-negligible.

\note{
However, due to their focus on contemporary processor architectures,
these works rely on an impenetrable kernel for recovery of a failed OS
instance.  The complexity of this kernel is comparable to that of a
small hypervisor. For example, OSIRIS~\cite{osiris} directs OS
recovery to a 29 KLOC reliable computing base (RCB)~\cite{rcb},
roughly twice the size of
modern
microkernels~\cite{Liedtke:sosp:l4:1995,l4fiasco,Klein+al:sosp:sel4:2009,asmussen:m3}.
}

Several other works have given early steps in the direction of the solutions
we advocate in this paper, minimizing the threat surface, or enforcing isolation.
Nohype
~\cite{Szefer:2011:EHA:2046707.2046754}
removes all
but a small kernel substrate from application cores, which run
functionality-rich OSs in virtual machines (VMs), reducing the threat surface.
\note{
... since hypervisor breaches now
have to go through redirected and remotely executed
hypercalls. However, although
necessary to raise the bar for adversaries, such intrusion prevention
techniques remain imperfect as a final solution.
}
%
Cap~\cite{needham:cap} and M3~\cite{asmussen:m3} exploit hardware
capability units and Hive~\cite{hive}
a bus-level firewall to isolate VMs at tile granularity.
However, although this avoids trusting tile-local kernel substrates for
isolation, their configuration interface, which is necessary to retain
flexible resource sharing, turns the configuring kernel into a
single point of failure. We address this problem in \midir.
\note{
We
follow Cap, M3 and Hive in relying on 
hardware to enforce privileges
and thereby confine faults to
tiles and the resources they are authorized to access, but advance
from these works by leaving clear
the need
to perform
any critical privilege and configuration changes in a consensual manner, to prevent
common-mode failures by fault propagation.
The concrete instance we
discuss is based on
capabilities, however, consensual update of other privilege mechanisms
achieves similar protection.

Capabilities are cryptographically-~\cite{amoeba},
kernel-~\cite{Hardy:1985:KA:858336.858337,eros,l4fiasco} or
hardware-protected tuples~\cite{needham:cap,asmussen:m3} comprised of at
least a pointer to an object (or service) and access rights
authorizing which operations owners of these capabilities may execute
on the object.
Posession of a capability is both necessary and sufficient to exercise
a granted access over an object. Consequently, as long as both
capability-enforcement and -reconfiguration are trustworthy, faults in
a component cannot propagate beyond the objects it can access, unless
other accessing components are faulty as well.
}

Cheri~\cite{Woodruff:2014:CCM:2665671.2665740} adds capability
protection on top of 
page-based protection, but includes the MMU and the OS page-table
management in the
reliable computing base (RCB), which means the former must be
trustworthy.
The concept behind \midir is independent of the protection model, not
being necessarily tied to e.g., capabilities.  
Also, establishing the fault containment domains at the granularity of
tiles, we are agnostic about the semantics and interplay of
tile-internal and/or core-level components, e.g., MMUs, memory
protection or page-table management.  Enforced by \bluedot, the
protection mechanisms are crafted at inter-tile level, emulating the
spacial isolation of distributed system nodes.

\note{
Cheri~\cite{Woodruff:2014:CCM:2665671.2665740} adds
capability protection on top of OS managed page-based protection, but
includes the MMU and the 
OS in each application's reliable computing base (RCB). 
%
%
We avoid this and 
establish tiles as fault containment domains by channeling
accesses to other tiles and to tile-external resources 
through the hardware capability unit 
\bluedot for privilege confirmation. In \midir, an instance of
\bluedot is located between each tile and its NoC
interface. Tile-external accesses include memory reads and writes
(e.g., issued by the tile's cores),
but also messages that are
interpreted as
IRQs
or IO. Therefore, by
constraining the capabilities of a tile, we
also confine how
faults in this tile may propagate.
}

Replication has been used before in closely-coupled systems, primarily
to tolerate accidental faults in cyber-physical systems (CPS), by
replicating controllers to form triple modular redundant (TMR) units,
or duplicated self-checking units. An example of the use of TMR in highly critical systems 
can be seen in the primary flight computers of Boeing 777's fly-by-wire (FBW) system~\cite{yeh1998triple}. In a similar context, a form of passive redundancy can also be seen in Airbus' dependability-oriented approach to FBW, where "hot spares" are used in case the active computer 
interrupts its activity~\cite{traverse2004airbus}.
The concept was extended to multi-phase tightly synchronous
message-passing protocols still in the CPS domain~\cite{mancini1986modular, kopetz2003time}.
The so-called 'Paxos'~\cite{schiper2014developing}, and 'Byzantine'~\cite{pbft} Fault-Tolerant
State-Machine Replication classes of protocols
 promote resilience to threats, respectively accidental, and both
 accidental and malicious, extending the concept to generic classes of
 applications, namely in loosely-coupled systems.
%
For example, Castro's seminal BFT-SMR protocol~\cite{pbft} masks the
actions of a minority of up to $f$ compromised replicas, by reaching a
majority voted consensus of $|Q|=2f+1$ out of $n=3f+1$ replicas.
Behind all the categories of techniques above is a baseline voting
mechanism amongst the values proposed by a pre-defined number of
replicated fault-independent components. 
\midir offers such baseline mechanism at a low enough level of
abstraction to serve essentially any replication-oriented application.

Architectural
hybridization~\cite{Verissimo:2006:TTW:1122480.1122497} (i.e.,
the inclusion of trusted-trustworthy components that follow a
differentiated fault model) allows reducing $n$ and $|Q|$ to $2f+1$ and
$f+1$, respectively~\cite{1353018,levin2009trinc,veronese2013efficient,cheapbft}.
%
The implementation of \bluedot, the \midir hybrid, draws from these
quorum reduction results, and further accelerates the BFT-SMR
protocol that \midir-enabled FIT microhypervisors use to coordinate
system-call execution (Sec.~\ref{sec:hypervisor}).

Paxos and BFT replication have been attempted as well inside
MPSoCs~\cite{bressoud:hypervisor_ft,kernel_paxos,paxos,romain,barrelfish}. However,
all these works were made under the assumption of a trusted low-level
kernel (e.g., hypervisor or platform manager), which obviously is a
single point of failure (SPoF). One of the key results of \midir lies
in the realization of the distributed system-on-a-chip (DSoC) vision,
which enables such replication management techniques in MPSoCs, whilst
removing the SPoF syndrome of the low-level kernel.

\note{
Under the assumption of a trusted low-level kernel (e.g., hypervisor
or platform manager), BFT-SMR inside MPSoCs becomes relatively
straightforward.
For example, Bressoud and Schneider~\cite{bressoud:hypervisor_ft}
pioneered hypervisor support for replica coordination, Esposito et
al.~\cite{kernel_paxos} implement a Paxos~\cite{paxos} kernel module
and D\"obel et al.~\cite{romain} microkernel-based replication.
Barrelfish~\cite{barrelfish,consensus-inside} replicates kernel state
across tiles and implements a two-phase commit-like protocol to update
this state.
Chun et al.~\cite{Chun:2008:DRS:1404014.1404038} identify
optimizations to BFT that are, however, not possible in the distributed
case.
Nonetheless, as we have discussed, today's reality calling for
removing that (non-substantiated) assumption, the solutions above are
plagued by the problem we set out to solve in this paper: threats
leveraging the SPoF syndrome of the low-level kernel.
}

\note{
We draw from these results to implement \bluedot, \midir's trustworthy
hardware component, to constrain fault propagation and allow for
correct privilege change in the presence of failed or compromised
tiles. We also leverage the same hardware components to accellerate
the BFT-SMR protocol that \midir-enabled FIT microhypervisors use to
coordinate system-call execution.
}

\note{
Other works discuss another important facet: fundamental conditions
and techniques to prevent exhaustion failure under persistent threats
(loss of the majority of correct replicas) to achieve
sustainability. Several techniques are proposed, such as proactive and
reactive rejuvenation and resilience, and diversification addressing
fault
independence~\cite{Sousa:2010:HAI:1749403.1749508,schneider:obfuscation,Larsen:2014:SAS:2650286.2650803,xu:randomization,garcia2019lazarus}.
Though not presented here for lack of space, \midir offers a
low-level rejuvenation protocol, and relies on diversity to avoid
common mode faults.
}

\section{Conclusions and Future Work}
\label{sec:conclusions}

We have introduced \midir, an architectural concept
which breaks new ground and opens promising avenues in the
applicability and resilience of manycore architectures (MPSoC).
%
Through minimalist mechanisms integrated in the MPSoC architecture,
\midir frees MPSoCs from the SPoF syndrome, fulfilling the vision of
\emph{distributed} systems-on-a-chip (DSoC).

In this paper, we show in particular that \midir-enabled DSoCs achieve
a quantum step towards off-the-shelf chip resilience, since these
mechanisms are generic enough to support, in-chip and with high
reliability, a large variety of the protection and redundancy
management techniques normally implemented in software at higher
layers in 'macro' systems.
%
%
%
\note{
\midir achieves a quantum step towards off-the-shelf chip resilience,
since these mechanisms are generic enough to support, in-chip and with
high reliability, a large variety of the protection and redundancy
management techniques normally implemented in software at higher
layers in 'macro' systems. Being anchored on simple hardware
extensions (capability registers and quorum-based voters - \bluedot)
staged at the tile-to-NoC interface, they may be easily incorporated
by chip manufacturers
in a non-intrusive way, preserving their architecture legacy.
}
To convincingly prove our point, we exemplified and
evaluated an implementation, over \midir, of the most complex version 
of our solution set: a Byzantine fault tolerant microhypervisor. 
We  have shown the
practicality of our concept, as having quite satisfying performance, 
since it outperforms the highly efficient MinBFT protocol by one 
order of magnitude.
%
The low overhead of our approach shows as well large
promise for future full hardware solutions.

Furthermore, \midir was intentionally designed as a non-intrusive
extension to current chip architectures, being anchored on simple and
self-contained hardware extensions.
Taken up by a hardware manufacturer or integrator, it allows a
backward compatible, non-fracturing evolution.  We hope that our
findings may be key to enhance general MPSoC architectures towards
distributed DSoCs and amongst other avenues, lead to next-generation
COTS resilient chips.
%
%

After this initial work, several questions remain to be answered,
namely on kernel design details,
rejuvenation and diversification for sustainability, and so forth,
which leave ample room for future work.

\clearpage

\bibliographystyle{IEEEtran}
\bibliography{midir}

\end{document}